\documentclass[]{aa}

\newcommand{\kms}{\ensuremath{\text{km}\,\text{s}^{-1}}}

\usepackage{graphicx}
\usepackage{xcolor}
\usepackage{txfonts}
\usepackage{amssymb}
\usepackage[]{hyperref}
\title{PM~1-322: new variable planetary nebula}

\author{
E.~Paunzen\inst{1}
          \and
K.~Bernhard\inst{2,3}
          \and
          J.~Budaj\inst{4}
          \and
          F.-J.~Hambsch\inst{2,5}
          \and
          S.~H{\"u}mmerich\inst{2,3}
          \and   
          D.~Jones\inst{6,7,8}
          \and
          J.~Krti\v{c}ka\inst{1}
          }

\institute{
Faculty of Science, Masaryk University, Department of Theoretical Physics and 
              Astrophysics, Kotl\'{a}\v{r}sk\'{a} 2, 611\,37 Brno,  Czech Republic
            \and
Bundesdeutsche Arbeitsgemeinschaft f{\"u}r Ver{\"a}nderliche Sterne e.V. (BAV), Berlin,
              Germany 
          \and
              American Association of Variable Star Observers (AAVSO), Cambridge, USA 
          \and
              Astronomical Institute, Slovak Academy of Sciences, 05960 Tatransk{\'a} Lomnica, 
              Slovak Republic
          \and
                    Vereniging Voor Sterrenkunde (VVS), Brugge, BE-8000, Belgium
          \and  
            Instituto de Astrof\'isica de Canarias, E-38205 La Laguna, Tenerife, Spain
            \and
            Departamento de Astrof\'isica, Universidad de La Laguna, E-38206 La Laguna, Tenerife, Spain
            \and
            Nordic Optical Telescope, Rambla Jos\'e Ana Fern\'andez P\'erez 7, 38711, Bre\~na Baja, Spain}

\date{Received ; accepted }
 
\abstract
{Spectra of planetary nebulae (PNe) are characterised by strong forbidden emission lines and often also by an infrared (IR) excess. A few PNe show dust obscuration events and/or harbour long-period binaries. Some post-asymptotic giant branch stars, symbiotic stars, or B[e] stars may feature similar characteristics. Recently, dust clouds eclipsing white dwarfs were also detected.}
{We report the discovery of an object with a very peculiar variability pattern that bears signatures compatible with the above-mentioned classes of objects. The object is ZTFJ201451.59+120353.4 and identifies with PM~1-322.}
{The object was discovered in Zwicky Transient Facility archival data and investigated with historical and newly obtained photometric and spectroscopic observations.}
{The ZTF $r$ and $g$ data show a one magnitude deep, eclipse-like event with a duration of about half a year that occurred in 2022. The variability pattern of the star is further characterised by several dimming events in the optical region that are accompanied by simultaneous brightenings in the red and IR regions. Apart from that, two fast eruption-like events were recorded in ZTF $r$ data. Archival data from WISE indicate long-term variability with a possible period of 6 or 12 yr. Our follow-up time series photometry reveals a stochastic short-term variability with an amplitude of about 0.1 mag on a timescale of about one hour.
The spectral energy distribution is dominated by IR radiation. Our high-resolution spectroscopy shows strong forbidden emission lines from highly ionised species and symmetric double-peaked emission in H$\alpha$, which is very different from what is seen in earlier spectra obtained in 2007.}
{Several explanatory scenarios are presented. Our most likely interpretation is that our target object involves a hot central star surrounded by gaseous and dusty disks, an extended nebula, and a possible companion star. Further observations are required to shed more light on the true nature of this enigmatic object.}

   
\begin{document}
\maketitle

\titlerunning{PM~1-322}
\authorrunning{Paunzen et al.}

\section{Introduction} \label{introduction}

Most planetary nebulae (PNe) are far from homogeneous spherical objects. On the contrary, a plethora of structural differences in PNe suggests that binary stars and their evolution 
may play a significant role in {\bf PNe} formation.
In a binary system, an asymptotic giant branch (AGB) star overflows the Roche lobe creating a common envelope (CE) engulfing the binary. The outcome is a short-period binary or a merger, and an ejected CE that forms the PN. 

Mass transfer in wide binary systems may also lead to the PN phase. AGB stars suffer huge mass loss through stellar wind, the velocity of which is comparable
to the orbital velocity. This results in an accumulation of material in the orbital plane. Long-period binaries were indeed confirmed in the cases of PN~G052.7+50.7, PN~LoTr5, and NGC~1514 \citep{vanwinckel14,jones17}.

A few central stars of PNe also show dust obscuration events, such as V651 Mon
\citep{kohoutek82,mendez82}, CPD-56$^\circ$8032 \citep{pollacco92}, or M 2-29 \citep{miszalski11}.
A recent review of PNe and the role of binaries can be found in \cite{jones17nat}.

The spectra of PNe are characterised by strong emission lines, most of which are forbidden lines originating from meta-stable levels of ionised species \citep{wesson16,lee13}. Such lines typically occur in vast rarefied gas regions irradiated by strong ultraviolet (UV) light.
However, similar conditions and lines can be found in post-AGB stars, B[e] stars, or symbiotic stars. 

Many post-AGB stars are long-period binaries and have dusty disks - the so-called Van Winckel objects \citep{vanwinckel03}.  However, most of these objects completely avoid the PN phase; they thus seem to follow a completely different evolutionary pathway, although there are many similarities with the wider binary central stars of PNe \citep{vanwinckel14}.

Symbiotic stars (SySts) are long-period binaries by definition. There are three distinct subclasses of these objects. One of them, D'-type, shows a very pronounced infrared (IR) excess peaking at 20$-$30 micron \citep{allen82}. It was suggested that they may in fact be compact PNe \citep{kenyon88}. Some SySts feature eclipses and strong forbidden emission lines \citep{skopal05}.

B[e] stars are also often characterised by strong double-peaked H$\alpha$ emission lines, narrow forbidden low-excitation emission lines ([Fe~\textsc{ii}], [OI]), and a strong IR excess \citep{zickgraf03}. Some of them may also be unresolved compact PNe \citep{lamers98}.

White dwarfs (WDs) are direct descendants of PNe. Some of them feature an IR excess and, recently, a few WDs were discovered that are eclipsed by dust clouds associated with exoasteroids \citep{vanderburg15,vanderburg20,vanderbosch20,manser19,gansicke19}.  

 This paper presents a study of the nebula PM~1-322, which bears many similarities to the above-mentioned objects. Its special properties were discovered by one of us (K.\,Bernhard) as a byproduct of the search for anti-phase variability at different wavelengths.
 Anti-phase variations in different photometric passbands are rarely observed and were previously considered an almost unique feature of a subgroup of $\alpha^2$ Canum Venaticorum (ACV) stars (see e.g. \citealt{2021A&A...656A.125F}). The search was carried out by a visual inspection of the $r$ and $g$ band light curves of suspected variable stars from the Zwicky Transient Facility (ZTF) Catalog of Periodic Variable Stars \citep{2020ApJS..249...18C}. In addition to several unpublished strictly periodic anti-phase ACV stars, there was also a single object (ZTFJ201451.59+120353.4) that showed anti-phase variability between the $r$ and $g$ band light curves but apparently no strict periodicity, which, however, is a characteristic of ACV variables. To the best of our knowledge, a similar behaviour has not yet been described in the literature, which is what initiated this study. 
 ZTFJ201451.59+120353.4 identifies with PM~1-322.
\begin{figure}[t]
\begin{center}
\includegraphics[width=0.45\textwidth]{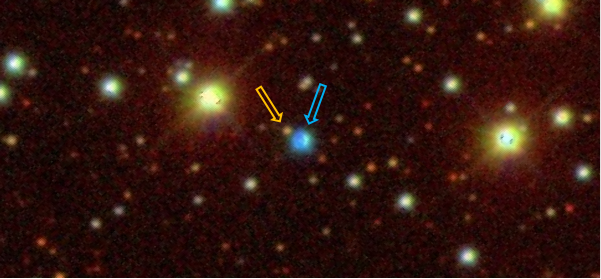}
\caption{Surroundings of PM~1-322, taken from the Sloan Digital Sky Survey \citep{2012ApJS..203...21A}. The blue arrow points to PM~1-322, while the orange arrow indicates $Gaia$ DR3 
1803129048700838016, which is about 4$\farcs$2 from PM~1-322.}
\label{sdss9}
\end{center}
\end{figure}

\section{An overview of PM~1-322 from the literature} \label{overview}

\citet{2005A&A...433..579P} were the first to report the discovery of PM~1-322 as a young PN on the basis of spectroscopic observations. Because of the high electron density and the
position in the line strengths 4363/H$\gamma$ versus 5007/H$\beta$ diagram (see Fig. 3 therein), the authors already discussed the possibility of the object being a symbiotic star. Their main argument against the symbiotic star hypothesis was the absence of any absorption line or TiO band in the spectra as well as no indication for an increase of the continuum towards the red, which would indicate the presence of a companion. Therefore, they favoured the interpretation that PM~1-322 is a young high-density PN.

High-resolution long-slit spectra were presented by \citet{2010PASA...27..199M}. The nebula was 
unresolved in their direct images. From the [\ion{N}{ii}] and [\ion{O}{iii}] single peak profiles, they derived
a systemic velocity of the nebula of +27.2\,\kms. The H$\alpha$ emission line showed a double-peaked profile with a centroid radial velocity in excellent agreement with the value listed above. The two emission peaks of the H$\alpha$ profile
were observed at $-$17.5 and +9\,\kms\ with respect to the
system velocity. The red peak was stronger than the blue one. Because of the large differences between the forbidden lines and
the H$\alpha$ profiles, \citet{2010PASA...27..199M} suggested a different origin for these spectral features. In particular, the relatively broad wings of the H$\alpha$ emission line were thought to possibly be caused by Rayleigh-Raman scattering in a very dense region close to the central star. The authors concluded that 
the spectral properties of PM~1-322 are remarkably similar to the properties of other PNe suspected to host a symbiotic central star \citep{2010PASA...27..129F}. 

\citet{2019ApJS..240...21A} classified PM~1-322 as a D'-type SySt. In these stars, both components, the giant cool star and the dust shells, contribute to the total spectral energy distribution (SED), which results in 
a nearly flat profile. The above-mentioned authors derived an effective temperature of the cool companion of 3811\,K
and a primary dust shell temperature of 722\,K. No \ion{O}{vi} 6830 line was detected. This emission feature was detected in about 50\% to 60\% of all Galactic SySts and is used as a diagnostic tool for their characterisation \citep{1989A&A...211L..31S,2019ApJS..240...21A}. It was identified as Raman scattering of the \ion{O}{vi} resonance doublet at 1032 and 1038\,\AA\ by neutral hydrogen.

\begin{table}
\caption{Data from $Gaia$ DR2 and DR3 for PM~1-322 and its neighbouring object $Gaia$ DR3 
1803129048700838016 (cf. Figure \ref{sdss9}).}
\label{table_Gaia}
\tiny
\begin{center}
\begin{tabular}{ccc}
\hline
\hline
& PM~1-322 & NBO \\
\hline
ID (DR2) & 1803129044411831680 & 1803129048700838016 \\
ID (DR3) & 1803129044411831680 & 1803129048700838016 \\
RA (DR2) & 303$\fdg$71499823 & 303$\fdg$71599019 \\
RA (DR3) & 303$\fdg$71499775 & 303$\fdg$71598999 \\
DEC (DR2) & +12$\fdg$06481596 & +12$\fdg$06548422 \\
DEC (DR3) &  +12$\fdg$06481521 & +12$\fdg$06548302 \\ 
$G$ (DR2) & 15.393$\pm$0.006\,mag & 19.521$\pm$0.004\,mag	\\
$G$ (DR3) & 15.410$\pm$0.006\,mag & 19.485$\pm$0.004\,mag	\\
$\mu_{\alpha\cos\delta}$ (DR2) & $-$4.789$\pm$0.111\,mas\,yr$^{-1}$ & $-$0.648$\pm$0.766\,mas\,yr$^{-1}$ \\
$\mu_{\alpha\cos\delta}$ (DR3) & $-$2.276$\pm$0.088\,mas\,yr$^{-1}$ & $-$1.150$\pm$0.344\,mas\,yr$^{-1}$	\\
$\mu_{\delta}$ (DR2) & $-$5.077$\pm$0.091\,mas\,yr$^{-1}$ & $-$8.761$\pm$0.765\,mas\,yr$^{-1}$ \\
$\mu_{\delta}$ (DR3) & $-$5.795$\pm$0.075\,mas\,yr$^{-1}$	& $-$8.275$\pm$0.358\,mas\,yr$^{-1}$ \\
$\pi$ (DR2) & 0.7107$\pm$0.0735\,mas & 0.5304$\pm$0.5439\,mas \\
$\pi$ (DR3) & 0.4801$\pm$0.0799\,mas & 0.7799$\pm$0.3689\,mas	\\
Dup (DR2) & 0 & 0 \\
Dup (DR3) & 0 & 0 \\
RUWE (DR3) & 3.022 & 1.153 \\
\hline
\end{tabular}
\end{center}                               
\end{table}

\begin{figure}[t]
\begin{center}
\includegraphics[width=0.45\textwidth]{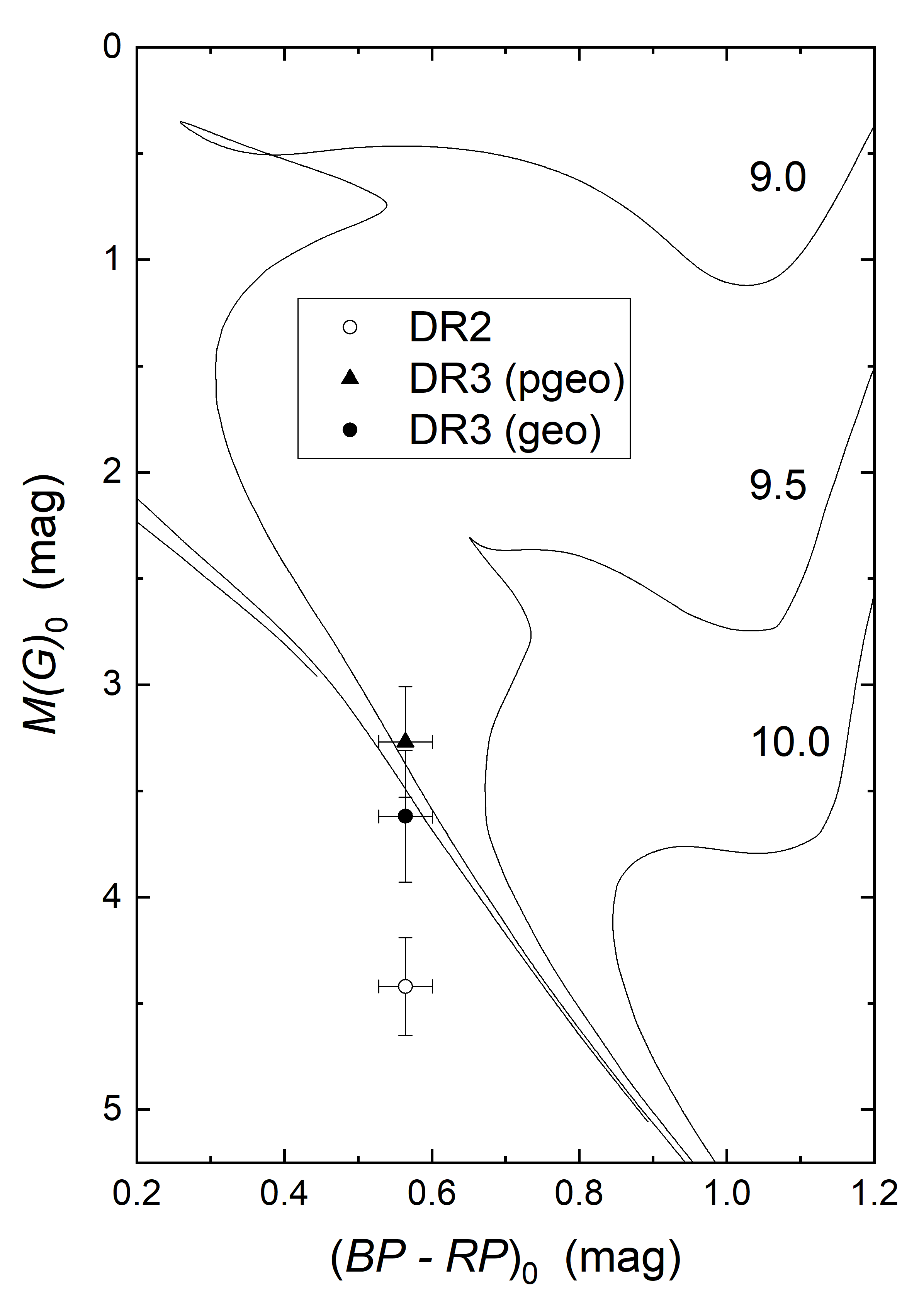}
\caption{Location of PM~1-322 in the $M$(G)$_\mathrm{0}$ versus ($BP-RP$)$_\mathrm{0}$ diagram, together with main-sequence PARSEC isochrones \citep{2012MNRAS.427..127B} for a solar metallicity of [Z]\,=\,0.0152. The numbers indicate $\log Age$ in yr. The different symbols mark the locations using three different approaches to convert the parallaxes from the $Gaia$ DR2 and DR3 data, as indicated in the in-set legend.}
\label{fig_CMD}
\end{center}
\end{figure}

\section{Results from $Gaia$} \label{results_gaia}

Because PM~1-322 is an extended source, one has to assess
the astrometric data of the $Gaia$ data releases carefully. The object $Gaia$ DR3 1803129048700838016 is about 4$\farcs$2 from the position of PM~1-322 (Fig. \ref{sdss9}). We searched the $Gaia$ data from DR2 \citep{2018A&A...616A...1G} and (E)DR3 \citep{2021A&A...649A...1G} for the corresponding entries of both objects.

In Table \ref{table_Gaia}, we summarise all relevant information from the two recent data releases. The data are all intrinsically consistent, with the exception of the parallax ($\pi$) for PM~1-322. Transforming the values into distances using the Bayesian approach by \citet{2018AJ....156...58B,2021AJ....161..147B}, we get a distance range from 1234 to 1521\,pc for the DR2 and from 1730 to 2299\,pc (geometric approach) as well as from 2095 to 2668\,pc (photogeometric approach) for the DR3 data. We therefore constructed a colour-magnitude diagram using the distance values of all three different approaches.

As next step, the interstellar reddening (absorption) was taken into account. For this, we relied on the three-dimensional reddening map of \citet{2019ApJ...887...93G}. The extinction in this direction is quite low and fairly constant over the listed distance range. The adopted absorption in $V$ is 0.356\,mag, which we transformed to the $Gaia$ DR2 photometric system using the coefficients listed in \citet{2018A&A...616A..10G}.

In Fig. \ref{fig_CMD}, we present the $M$(G)$_\mathrm{0}$ versus ($BP-RP$)$_\mathrm{0}$ diagram together with the main-sequence PARSEC isochrones \citep{2012MNRAS.427..127B} for a solar metallicity of [Z]\,=\,0.0152. The dereddened colour suggests an effective temperature of about 6600\,K. The distance from the $Gaia$ DR2 data would place the object significantly below the zero-age main sequence while the values from DR3 would place it at
or close to it.

Also included in Table \ref{table_Gaia} are the duplicity flag `Dup' and the Renormalised Unit Weight Error (RUWE). RUWE is expected to be around 1.0 for sources where the single-star model provides a good fit 
to the astrometric observations. Values significantly larger than 1.0 could indicate that the 
source is non-single or otherwise problematic for the astrometric 
solution\footnote{\url{https://www.cosmos.esa.int/web/gaia/public-dpac-documents}}.
The large RUWE value for PM~1-322 could be caused by its extended nature.

\section{Spectral energy distribution}
\label{sed}

Although the $Gaia$ filters are broad, it is possible that the measured fluxes are affected by variability or strong emission lines. In particular, the fluxes in the $r$ filter might be enhanced by strong H$\alpha$ emission, which would result in making PM~1-322 appear redder and cooler. The overall SED contains more information on the nature of the star.

\begin{figure}[ht]
\begin{center}
\includegraphics[width=0.48\textwidth]{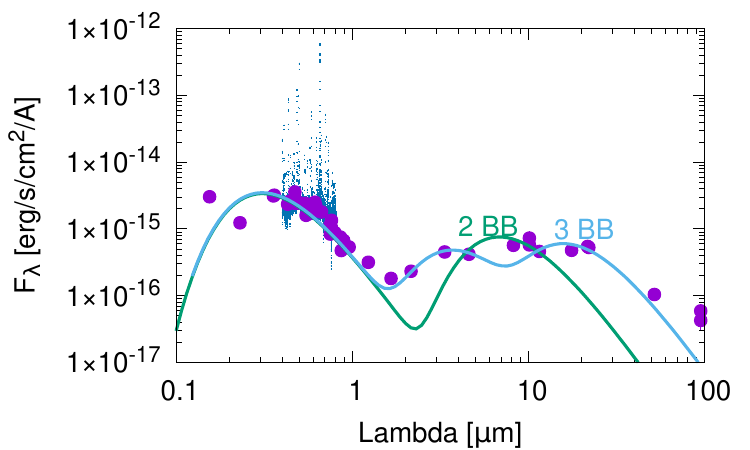}
\includegraphics[width=0.48\textwidth]{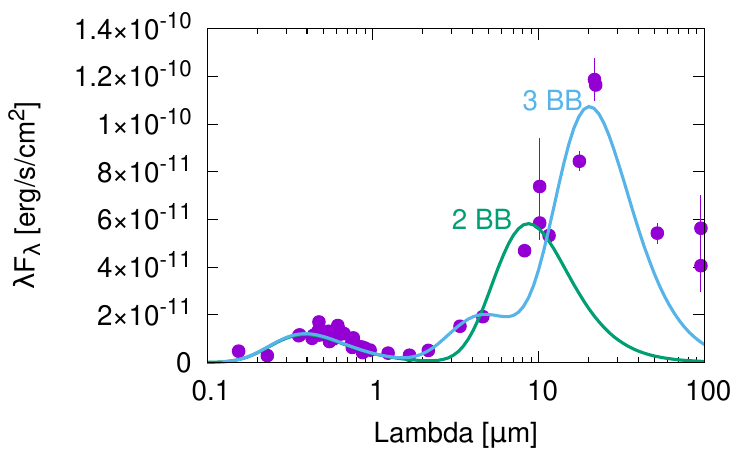}
\caption{SED of PM~1-322. Top: representation in log-log scale that clearly indicates an IR excess. Bottom: $\lambda F_{\lambda}$ representation that demonstrates that most of the energy is in fact radiated at IR wavelengths. Observations are indicated by the purple circles with error-bars. The green and blue lines denote, respectively, a simple two black-body fit and a three black-body fit to the data. The small dark-blue points in the upper plot represent a smoothed flux calibrated spectrum from Section~\ref{spectrum}.}
\label{fig:sed}
\end{center}
\end{figure}

The SED of our target object was constructed using the VOSA tool \citep{bayo08} and is shown in the top panel of Fig. \ref{fig:sed}. Only broad-band photometry was considered. The figure also includes our flux-calibrated spectrum, which is described in more detail in Section~\ref{spectrum}. This high-resolution spectrum was smoothed with a running window that was 2 \AA\ wide to enhance its signal-to-noise ratio and decrease the number of points. The spectrum contains strong emission lines but its continuum agrees well with the photometry. This indicates that the continuum is a major contributor to the flux at these broad-band photometric filters. A more detailed assessment of the contribution of the strongest emission lines to the flux at different filters is presented in Section~\ref{arch}. Typically, these lines contribute less than 30\% to the flux in these broad-band filters. However, variability of the source can affect the SED as well. As we see in Section~\ref{arch}, the star may experience brightness fluctuations by about 1 magnitude.  However, in the optical region, for most of the time, the amplitude of the variability is less than 0.3 mag.

The observations indicate strong flux in the UV and optical regions and a significant IR excess. However, the SED is not that of a typical symbiotic star in quiescence
\citep{skopal05}, which is dominated by the red giant star. The scatter is larger than the error bars, which is due to the variability of the source and its emission lines. For this reason, it is difficult to estimate the interstellar extinction and justify approximating the SED with stellar atmosphere models that feature absorption lines.
That is why, as a first approximation, we ignore the extinction\footnote{We tried to optimise and determine the reddening parameters but this did not lead to any improvement in the fit of the scattered SED values.} and fit the observed $F_{\lambda}$ fluxes using a simple model with two black bodies. The flux calibrated spectrum was not used in this fitting procedure.

Assuming a distance of 2083 pc, the temperatures and radii of the two black bodies are
$T_1=9400\,K, R_1=0.56\,R_{\odot}$ and $T_2=400\,K, R_2=600\,R_{\odot}$. The fit is also shown in Fig. \ref{fig:sed}.
We caution that although we use non-transparent spherical black-bodies, the real objects may be quite different. They do not have to be stars and they do not even have to be optically thick. For example, the first component could easily be an accretion disk hiding a much hotter central star (c.f. also Sections \ref{arch} and \ref{spectrum}).   

If there is a significant extinction, the temperature of the first (hotter) component will be higher.
Consequently, if the temperature is higher, the radius of the first component will be smaller. In either case, the radius of this component is much smaller than the radius of a star on the main sequence at such temperatures. It would rather correspond to a subdwarf B-type (sdB) star or the pseudo-photosphere of a white dwarf surrounded by circumstellar material.

If the secondary component is a star, its radius implies a supergiant. However, the temperature of this component is so low that it it obvious that the excess is caused by a dust cloud. Nevertheless, it is still conceivable that a star is embedded in the dust cloud, similar to what is observed in D'-type SySts. A similar infrared excess is also often observed in compact PNe. !!!

The log-log representation of the SED shown in the top panel of Fig.\ref{fig:sed} can be very misleading, which is why we also present the $\lambda F_{\lambda}$ versus $\log \lambda$
representation in the bottom panel. This kind of plot is better suited for assessing the actual SED. It demonstrates that most of the energy is actually radiated at IR wavelengths within the 7$-$40 micron interval. Assuming that a single hot star is responsible for heating the dust, its light emitted towards the observer has to be significantly attenuated, absorbed (e.g. by a dust shell or an edge-on disk), and reradiated in the IR region. This rules out a single star with an inclined dusty disk or ring configuration. This scenario also indicates that a simple two component black-body model completely fails to account for most of the energy coming from the star.

The situation improves if we add a third black body to the solution, which leads to the derivation of the following parameters:
$T_1=9400\,K, R_1=0.56\,R_{\odot}, T_2=800\,K, R_2=100\,R_{\odot}$, and $T_3=180\,K, R_3=4600\,R_{\odot}$, respectively. The fit is also shown in Figure \ref{fig:sed}. Since the third body radiates a lot of energy in the above mentioned 7$-$40 mic interval, it could be observed with the James Webb Space Telescope \citep{jwst06}. The angular radius of the third body on the sky would be about 10 mas, which indicates that such an optically-thick dust cloud will not be resolved. In case the cloud is optically thin, its radius might be much larger. However, optically-thin clouds could also radiate a lot of energy at these wavelengths due to the strong opacity of silicates, which would shrink the radius necessary to produce the observed amount of energy. Clearly, a more sophisticated model has to be used to account for the observed IR excess.

\section{Long-term variability} 
\label{arch}

\begin{figure}[ht]
\begin{center}
\includegraphics[width=0.48\textwidth]{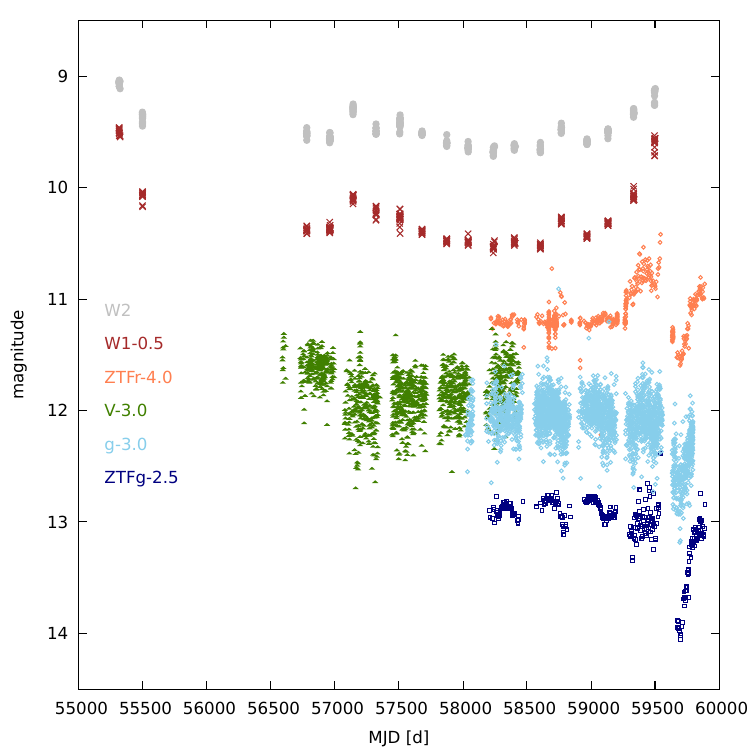}
\caption{Historical light curves of our target object in several filters. WISE/NEOWISE W2 filter: full grey circles, W1 filter: brown crosses. ASAS-SN $V$ filter: open green triangles, $g$ filter: sky-blue diamonds. ZTF $r$ filter: open orange circles, $g$ filter: open blue squares. Observations in some filters were offset by a constant value for clarity, as indicated in the plot by a number in magnitudes.}
\label{fig:arch}
\end{center}
\end{figure}

Several archives were searched for historic photomeric data on our target object. The ZTF \citep{bellm19} has observed this star in two filters (ZTF $g$ and ZTF $r$) for more than 4 years. The WISE \citep{wright10} and NEOWISE \citep{mainzer11} databases contain 12 yr of observations in the IR region
(W1 and W2 filters). Similarly, the All-Sky Automated Survey for Supernovae (ASAS-SN) contains 9 years of observations in the $V$ and $g$ filters \citep{shappe14,kochanek17}. For our analysis, we removed low-quality observations that only represent upper limits. The resulting collection of measurements is shown in Figure \ref{fig:arch}.

Two episodes of brightening by about one magnitude are observed in the W1 region, which are separated by 4500 days (12 yr). These are also seen in the W2 filter, albeit with reduced amplitudes. The most recent brightening was also captured in the ZTF $r$ filter. In between these two major events, there is a smaller brightening event at around MJD\,57200 that is noticeable in both IR filters.

\begin{figure}[ht]
\begin{center}
\includegraphics[width=0.48\textwidth]{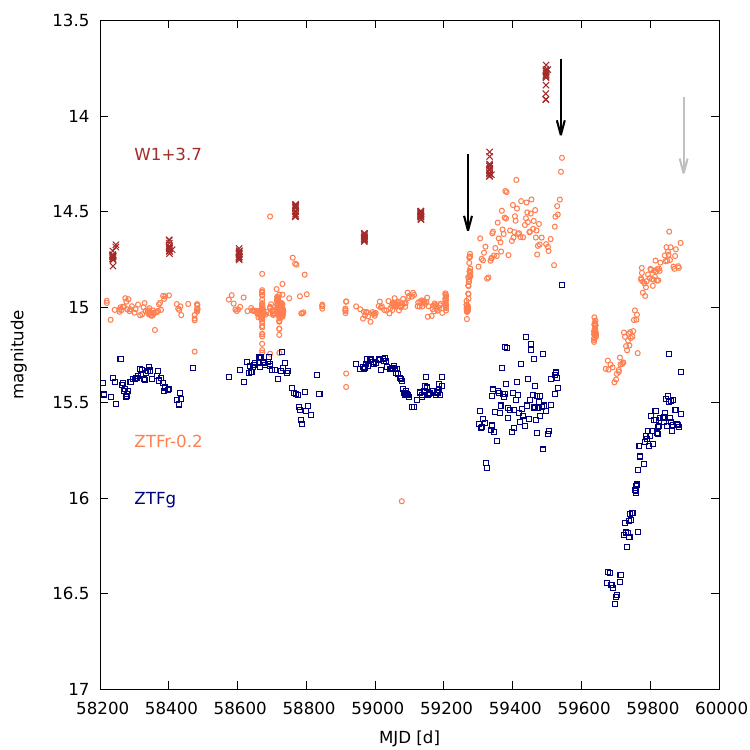}
\caption{Zoom-in on the light-curve evolution during the last four years of the ZTF coverage. WISE/NEOWISE W1 filter: brown crosses. ZTF $r$ filter: open orange circles, $g$ filter: open blue squares. Black arrows mark sudden brightening events. The grey arrow indicates the date of our spectroscopic observations. Observations in some filters were offset by a constant value for clarity, as indicated in the plot by a number in magnitudes.}
\label{fig:ztf}
\end{center}
\end{figure}

This kind of behaviour, however, is not observed in the optical region in the $V$ and $g$ filters, where there is a reverse trend with dimming events instead of brightenings. This is best illustrated in Figure \ref{fig:ztf}, which zooms in on the light curve evolution during the last four years of the ZTF coverage. A relatively quiet period during MJD\,58200$-$59200 indicates little dimming events in ZTF $g$ that are accompanied by small brightenings in ZTF $r$. A sudden (week-long) eruption-like event is observed in ZTF $r$ at MJD\,59270, which is then followed by additional activity and brightening until MJD\,59530, when another, even stronger (month-long) eruption-like event occurred. After this event, the brightness dropped by about one magnitude and we refer to this drop as `an eclipse'. Its duration is about half a year. The behaviour in the ZTF $g$ filter also shows enhanced activity in between the eruption-like events but there is no brightening. During the eclipse, however, the brightness also decreases in ZTF $g$ and the drop is significantly more pronounced than in the ZTF $r$ filter.

To explore the nature of this unusual variability, we calculated the synthetic photometry in ZTF $r$ and ZTF $g$ filters using our flux calibrated spectrum described in more detail in Section~\ref{spectrum}. Filter transmission curves and zero points were taken from the SVO filter profile service \citep{rodrigo12}. We obtained the following magnitudes: ZTF$g=15.43$, ZTF$r=14.53$. Then we removed the three strongest emission lines: H$\beta$, [OIII]5007, and H$\alpha$, recalculated the fluxes and magnitudes, and obtained ZTF$g=15.65$, ZTF$r=14.93$.
It shows that the emission lines contribute about 19\% and 31\% of flux in those filters, respectively. It also shows that any variability with an amplitude less than 0.2 mag in the $g$ filter and 0.4 mag in the $r$ filter, such as the eclipse, are unlikely to be caused by variability in emission lines or by the eclipse of a source of `pure' emission lines.
     
\begin{figure*}[h]
\centerline{
\includegraphics[width=0.48\textwidth]{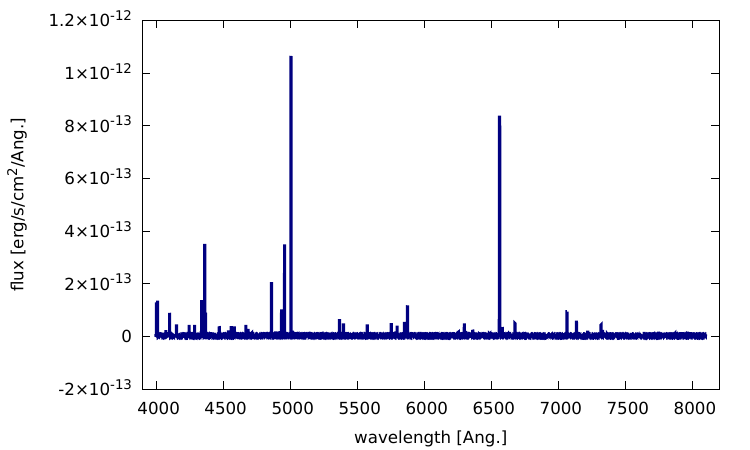}
\includegraphics[width=0.48\textwidth]{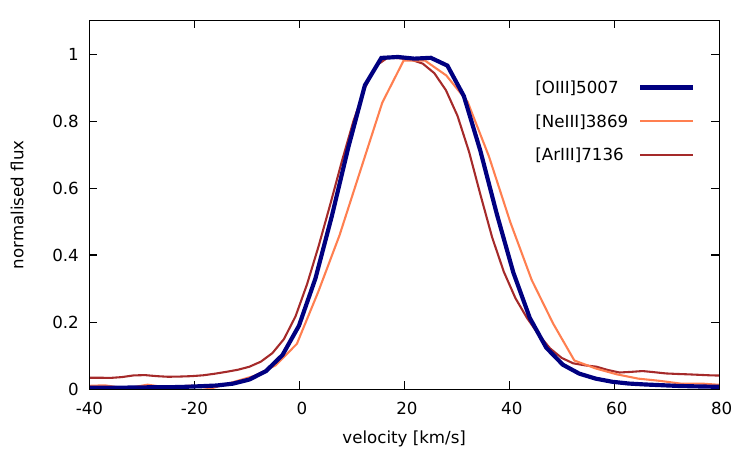}}
\centerline{
\includegraphics[width=0.48\textwidth]{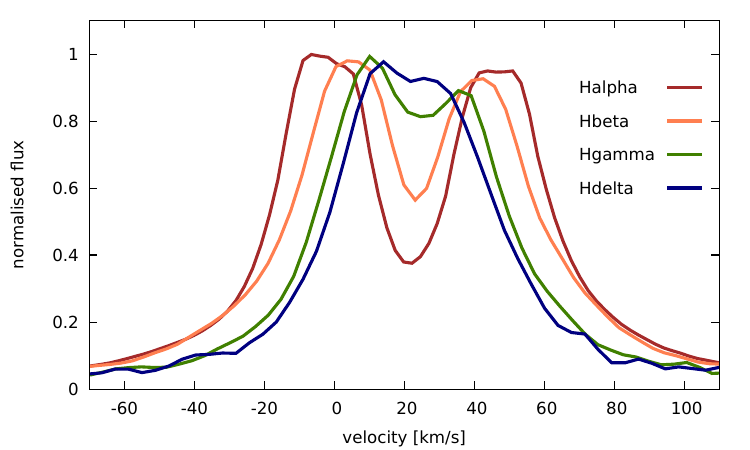}
\includegraphics[width=0.48\textwidth]{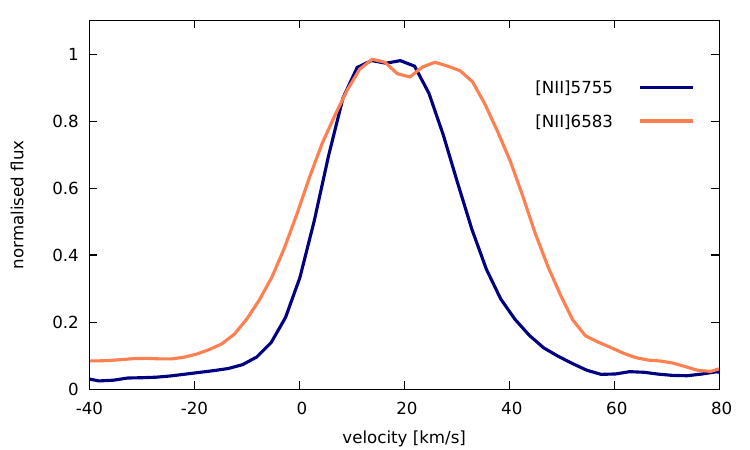}}
\caption{FIES spectrum and line-profiles. Top-left: complete FIES spectrum of PM~1-322. Top-right: [O~\textsc{iii}] 5007 line showing a flat-topped profile (the profile of [Ne~\textsc{iii}] 3869 is shown for comparison). Bottom-left: Balmer lines from lower levels develop a double-peaked profile. Bottom-right: [N~\textsc{ii}] lines from lower levels also exhibit a double-peaked profile. Line profiles were normalised to unity.}
\label{fig:allsp}
\end{figure*}

\section{Our follow-up observations}

\subsection{Spectroscopy} 
\label{spectrum}

PM~1-322 was observed with the FIES instrument \citep{2014AN....335...41T} on the 2.56-m Nordic Optical Telescope \cite[NOT;][]{2010ASSP...14..211D} on the night of November 14, 2022.  FIES was employed in low-resolution mode (fibre diameter of 2.5\arcsec{}) providing a reciprocal resolution of approximately 25000 from 3700--9000~\AA{}.  Three consecutive exposures of 1200s were obtained, followed by a single 300s exposure of the standard star BD+28 4211 \citep{stone77}. The airmass during the observations was approximately 1.4, as such the atmospheric dispersion corrector (ADC) was not used (based on commissioning tests, the ADC is recommended only for observations at airmasses $\gtrsim$ 1.5).
The data were reduced using the FIEStool pipeline\footnote{http://www.not.iac.es/instruments/fies/fiestool/FIEStool.html}. The flux calibration is only approximate given the fiber-fed nature of the observations. 

We note that our observations sample a different part of the sky compared to the previous observations of \cite{2005A&A...433..579P} or \cite{2010PASA...27..199M}, which used slits with a width of 4" and 1.6", respectively. However, since the nebula was spatially unresolved in H$\alpha$, [\ion{N}{II}], and [\ion{O}{III}] images with a resolution of about 1.8" \citep{2005A&A...433..579P}, it is unlikely that the slit width of 1.6", 4", or our aperture of 2.5" affects the spectral line profiles significantly.

The whole spectrum is shown in the top-left panel of Figure \ref{fig:allsp}. It consists of numerous strong (mostly forbidden) emission lines and a very weak continuum, which is not strong enough to permit equivalent width measurements or detect any absorption lines. Its typical S/N is about 5.
Most of the emission lines have nice symmetric profiles. Some lines, such as [O~\textsc{iii}] 4959, 5007, have curious flat peaks (see the top-right panel of Figure \ref{fig:allsp} for a comparison with [Ne~\textsc{iii}] 3869), which is not caused by saturation of the CCD detector. These lines are fairly narrow with a full width at half maxima of about 30 km/s, which is still above the limit dictated by the spectral resolution. Radial velocities derived from individual lines give a consistent value of about 
$22.1\pm 0.5$ km/s. This is significantly less than the value of 27.2 km/s estimated by \cite{2010PASA...27..199M}. However, we note that one has to keep in mind that our aperture was slightly different than that used by the aforementioned authors.

The Balmer series shows a very interesting behaviour. Lines originating from higher levels are narrower with a single peak. Lines from progressively lower levels are broader and develop a symmetric double-peaked profile (see the bottom-left panel of Figure \ref{fig:allsp}). H$\alpha$ shows a central depression deeper than the half of the maximum and wings that can be traced up to 400 km/s. Radial velocities of the blue emission, central absorption, and red emission are -2.3, 22.3, 46.3 km/s, respectively. This H$\alpha$ profile is very different from that observed in 2007 by \cite{2010PASA...27..199M}, in which the red emission was much stronger than the blue emission. Again, we note that our aperture was slightly different from the one used by these authors. The same pattern is observed in the nitrogen lines [N~\textsc{ii}] 5755, 6583 (Figure \ref{fig:allsp}, bottom-right panel). The first line originates from a higher level and is narrower while the second line is broader and double-peaked. The trend of narrower line widths for lines originating from higher Balmer lines is opposite compared to the dependence of peak separation on the principal quantum number of upper level observed in classical Be stars \citep{1988A&A...189..147H,2004A&A...419..607S}.

A possible interpretation of these observations is that the central star is embedded in a nebula and an almost edge-on disk. Assuming this configuration, one could speculate and argue along the following lines. Lines from the lower levels originate from the inner disk, that is a denser material in the orbital plane which is subject to Keplerian motion. The half-separation of the emission peaks in the H$\alpha$ profile is mainly sensitive to the radius of the disk and to its inclination. Assuming that the inclination is close to edge-on (see Section \ref{interp}) and the half-separation of the emission peaks is about 24 km/s, the radius where H$\alpha$ is produced would be about 1 au (assuming a central star slightly less massive than our Sun). The reason why low excitation Balmer lines originate from the inner disk could be that they are more opaque. There is no mutual radial velocity component between the disk material orbiting on circular orbits and the nucleus. Consequently, in the line core, the disk becomes opaque in the radial direction easily. The outer disk suffers from lower irradiation and, hence, less radiative excitation. An observer looking at the edge-on disk will not see very deep into the disk in the core of H$\alpha$. However, the blue and red wings of the line may be desaturated by the velocity gradient along the line-of-sight in the inner disk. Hence, in the line wings, the observer sees deeper and hotter regions of the inner disk with higher velocities. The denser material in the inner disk might also imply a stronger collisional excitation of lower levels.

Lines from the higher levels and forbidden lines would originate from either larger distances from the star, or from well above the disk plane, or from distant polar regions of the nebula where the densities are much lower. Assuming there is a hot star at the centre, these low density regions would be subject to considerably stronger UV irradiation since they would not be shielded by the disk. Such radiative excitation and ionisation would give rise to single-peaked forbidden emission lines from highly-ionised species. 

In the spectrum, we identified lines from H~\textsc{i}, He~\textsc{i}, He~\textsc{ii} (25\,eV), [N~\textsc{ii}], N~\textsc{iii} (30\,eV), [O~\textsc{i}, ~\textsc{ii}, ~\textsc{iii}] (35\,eV), O~\textsc{ii}, [Ne~\textsc{iii}] (41\,eV), [S~\textsc{ii}, ~\textsc{iii}] (23\,eV), [Cl~\textsc{iv}] (40\,eV), [Ar~\textsc{iii}, ~\textsc{iv}] (41\,eV), and [K~\textsc{iv}] (46\,eV). The numbers in parentheses indicate the ionisation potential in eV required to reach that particular ionisation state. This clearly supports the idea of a hot central star. Only the presence of [O~\textsc{iii}] requires temperatures above 25\,000\,K \citep{kwok00}. Following this scenario, the temperature of our hottest black body from Figure \ref{fig:sed}
would be slightly lower but an additional black body much hotter than 25\,000\,K (possibly shielded by the disk) would need to be introduced. This might also explain why the data point at the shortest wavelength in Figure \ref{fig:sed} exhibits a significantly increased flux compared to the models presented.

The explanation offered above may not be the only possible model. The double-peaked and flat-topped profiles might also originate from bipolar outflows or radially expanding latitude-dependent winds with dusty disks as seen in B[e]-type stars \citep{zickgraf03}. Some planetary nebula show a so-called Wilson effect \citep{wilson50}, which is when the separation between the blue and red peaks of lines of highly ionised species is smaller than that for less ionised ones. It is attributed to radially expanding envelopes.

\begin{table}[ht]
\caption{
Fluxes in some spectral lines integrated over the line profile relative to H$\beta$ in \%.
Fluxes in H$\beta$ in $erg/s/cm^2$ are given at the bottom together with some line ratios.
First column contains the line identification. In the second column are our measurements not corrected for dust extinction. The third column lists the measurements by \cite{2005A&A...433..579P} for comparison. }
\label{fluxes}
\centering
\begin{tabular}{lrr}
\hline\hline
 Line                   &This work& P\&M      \\
\hline
 [\ion{S}{II}] 4069     &    1.57 &     3.7  \\
 H$\gamma$ 4340         &   46.32 &    47.1  \\
 $[\ion{O}{III}]$ 4363  &   64.71 &   118.2  \\
 $[\ion{O}{III}]$ 4959  &   70.96 &   201.6  \\
 H$\beta$ 4861          &  100.00 &   100.0  \\
 $[\ion{O}{III}]$ 5007  &  222.79 &   612.9  \\
 \ion{He}{I} 5047       &    0.64 &    21.0  \\
 $[\ion{N}{II}]$ 5755   &   10.77 &    11.3  \\
 \ion{He}{I} 5876       &   48.16 &    28.3  \\
 $[\ion{K}{IV}]$ 6102   &    0.41 &   -      \\
 $[\ion{S}{III}]$ 6312  &    4.56 &     5.7  \\     
 H$\alpha$ 6563         &  595.59 &   393.2  \\ 
 $[\ion{N}{II}]$ 6583   &   11.62 &    20.2  \\
 \ion{He}{I} 6678       &   21.54 &     8.6  \\
 $[\ion{Ar}{III}]$ 7135 &   16.95 &    27.7  \\
 $[\ion{Ar}{IV}]$ 7171  &    1.27 &     1.8  \\
 $[\ion{Ar}{IV}]$ 7237  &    1.05 &     2.1  \\
 $[\ion{Cl}{IV}]$ 8046  &    0.53 &     1.5  \\
 \hline
 H$\beta$ $[erg/s/cm^2]$& $2.72\,10^{-13}$ & $2.5\,10^{-13}$  \\    
 4363/H$\gamma$         & 1.40    &  2.51    \\
 5007/H$\beta$          & 2.23    &  6.13    \\
\hline
\end{tabular}
\end{table}

In the flux calibrated spectra, we measured the fluxes of some of the most important spectral lines. They are given in Table \ref{fluxes} and compared with those from \cite{2005A&A...433..579P}. These are observed values not corrected for dust extinction or reddening. The most apparent change is that the forbidden oxygen lines, mainly [\ion{O}{III}] 5007, are now significantly weaker than before. H$\alpha$ got slightly stronger and is now the strongest line in the optical spectrum. Although its central intensity is smaller than that of 
[\ion{O}{III}] 5007 when plotted per unit of wavelength (see Fig. ~\ref{fig:allsp}), it carries most of the energy. \ion{He}{I} lines are also stronger now, except \ion{He}{I} 5047, but this is likely due to a misprint in the previous value. This indicates that densities and collisions in the environment have increased.
It is very convenient to use the 4363/H$\gamma$ and 5007/H$\beta$ line ratios, which are not sensitive to dust reddening. Both line ratios have decreased significantly, which places PM~1-322 deeper into the realm of the SySts (see Fig.~3 of \cite{2005A&A...433..579P}).

\subsection{Photometry} 

\begin{figure}[ht]
\begin{center}
\includegraphics[width=0.4\textwidth]{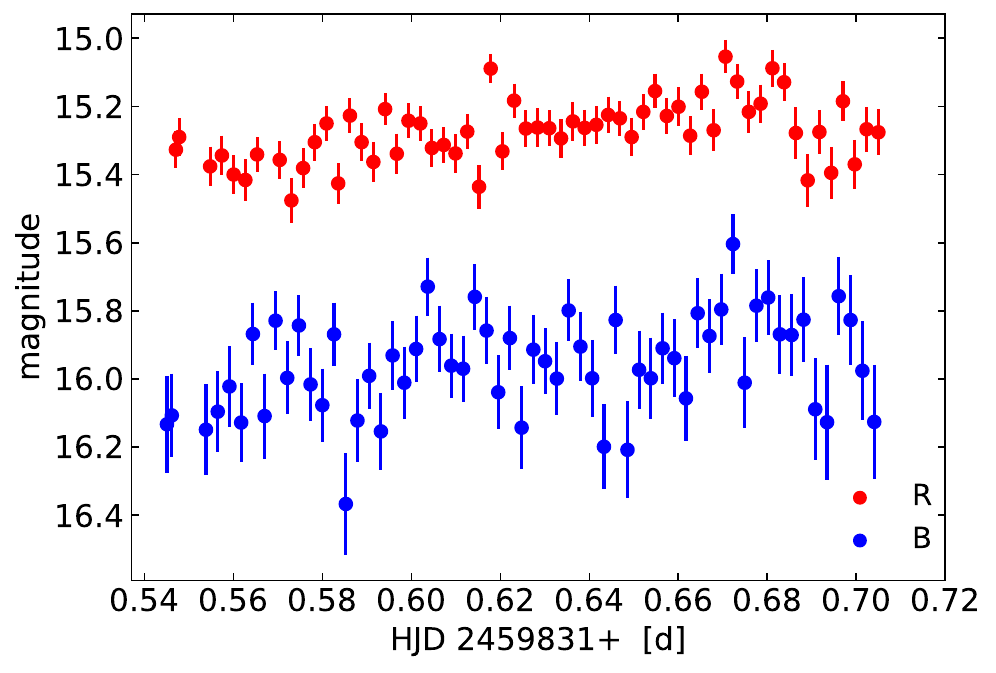}
\includegraphics[width=0.4\textwidth]{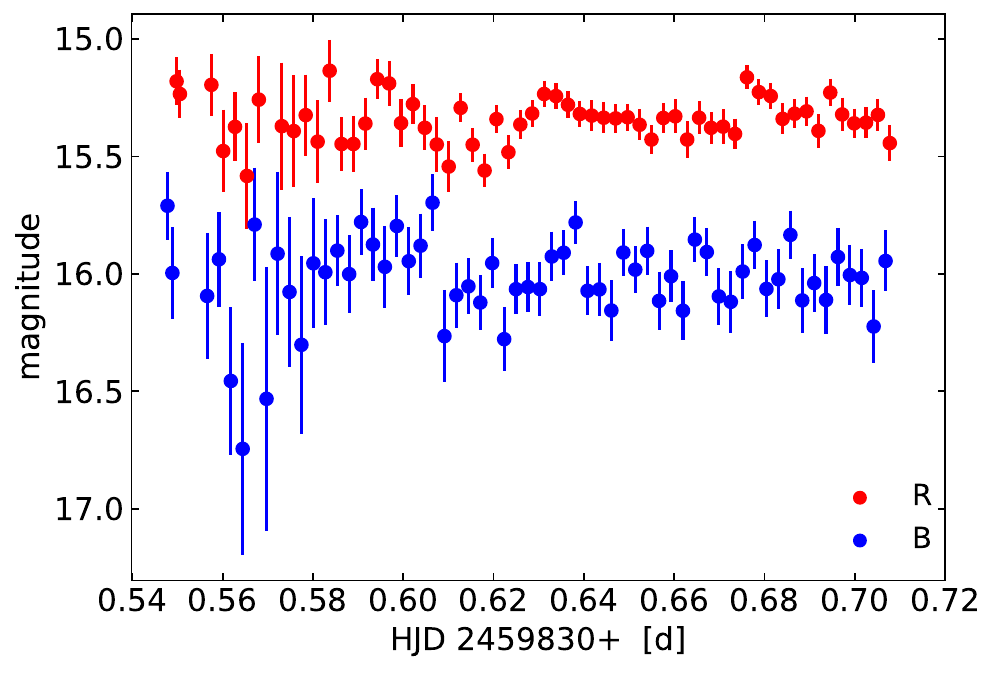}
\includegraphics[width=0.4\textwidth]{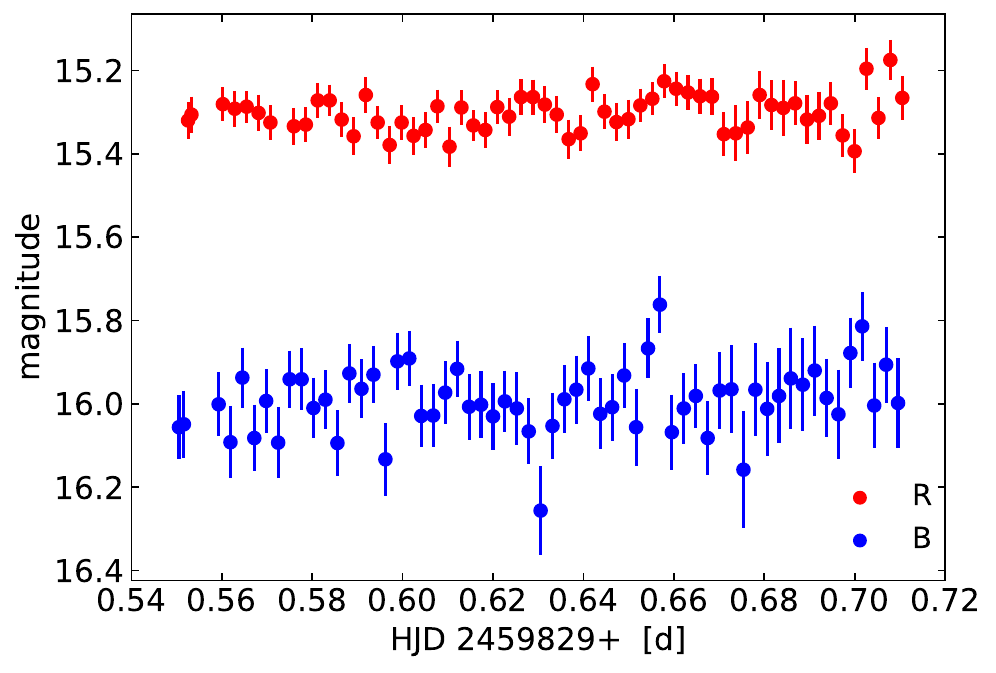}
\includegraphics[width=0.4\textwidth]{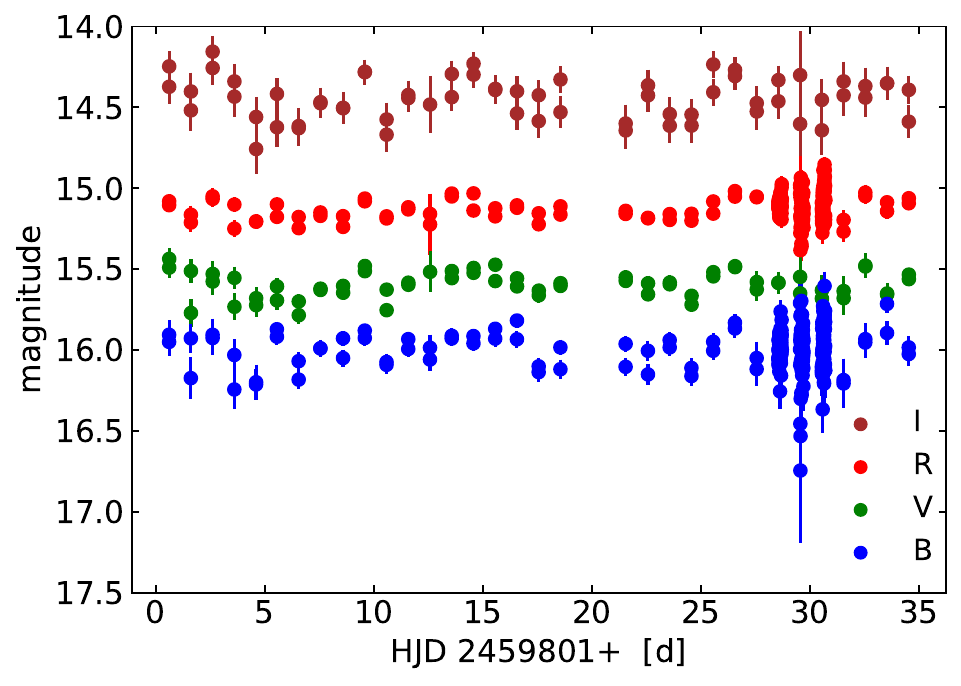}
\caption{ROAD times series photometry. Top three panels: $BR{_C}$ observations acquired on September 7 to 9. Bottom panel: the complete set of $BVR{_C}I{_C}$ observations.}
\label{fig:br}
\end{center}
\end{figure}

Follow-up observations were performed at the Remote Observatory Atacama Desert \citep[ROAD;][]{hambsch12}. The observations were acquired through Astrodon Photometric filters with an Orion Optics, UK Optimised Dall Kirkham 40 cm f/6.8 telescope and a FLI 16803 CCD camera. The field of view of the camera is 47 × 47 arcmin${^2}$. Each data set consists of pairs of exposures with 90\,s ($B$) or 60\,s ($V$, $R{_C}$, and $I{_C}$). Twilight sky-flat images were used for flat-field corrections. The observations covered the time span from August 9, 2022 to September 12, 2022.
On September 7 to 9, time series in $B$ and $R{_C}$ were taken over a period of about 3.6 hours each, which are depicted in Figure \ref{fig:br}. 

These data indicate variability with an amplitude of about 0.1\,mag on timescales as short as one hour. The finite speed of light therefore puts an important constraint on the scale of the objects involved in this kind of variability, which have to be smaller than a few au. We cannot, however, distinguish whether this variability originates from the variability in the continuum or from the emission lines.

\begin{figure}[ht]
\begin{center}
\includegraphics[width=0.48\textwidth]{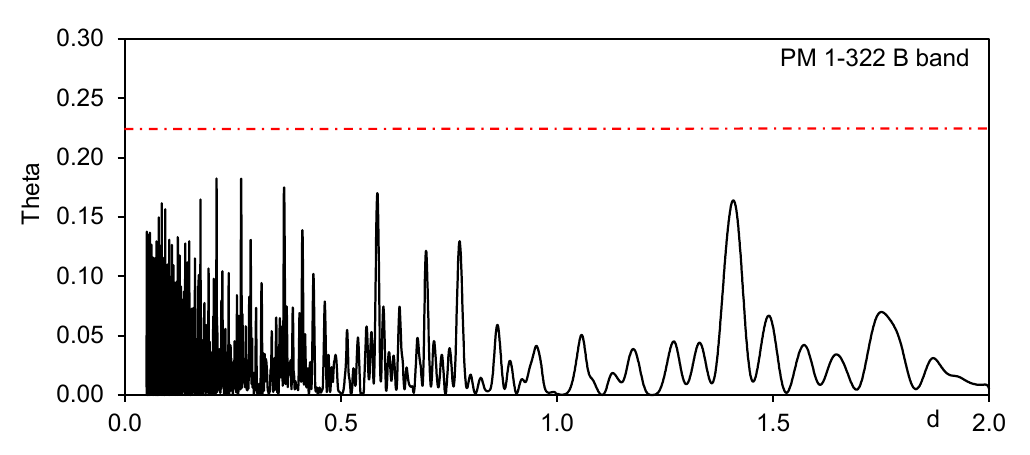}
\includegraphics[width=0.48\textwidth]{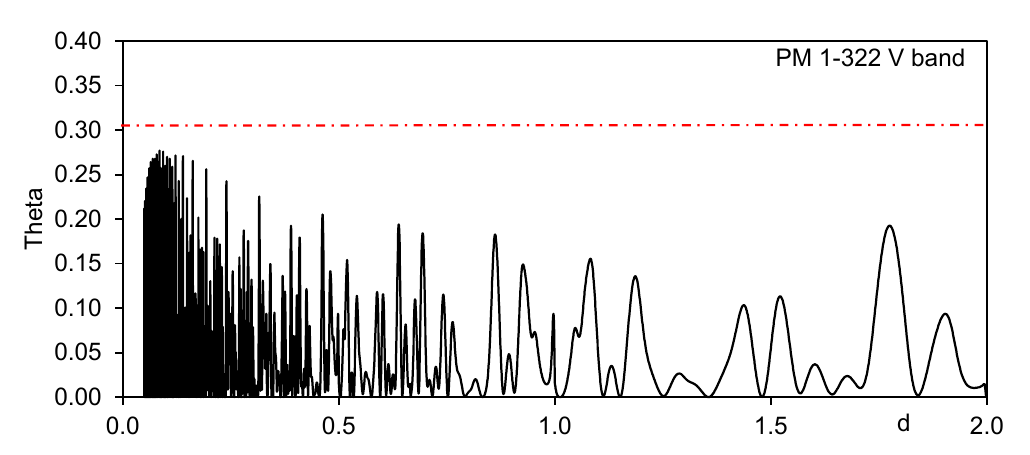}
\includegraphics[width=0.48\textwidth]{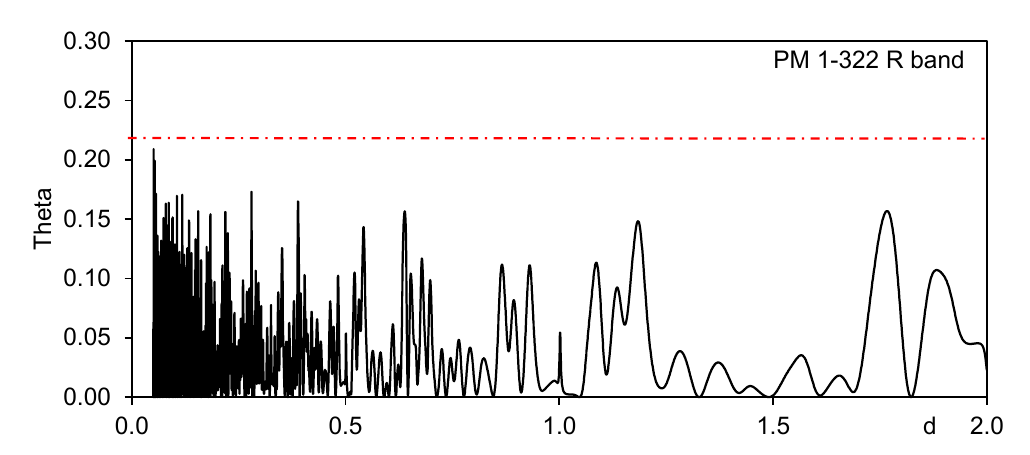}
\includegraphics[width=0.48\textwidth]{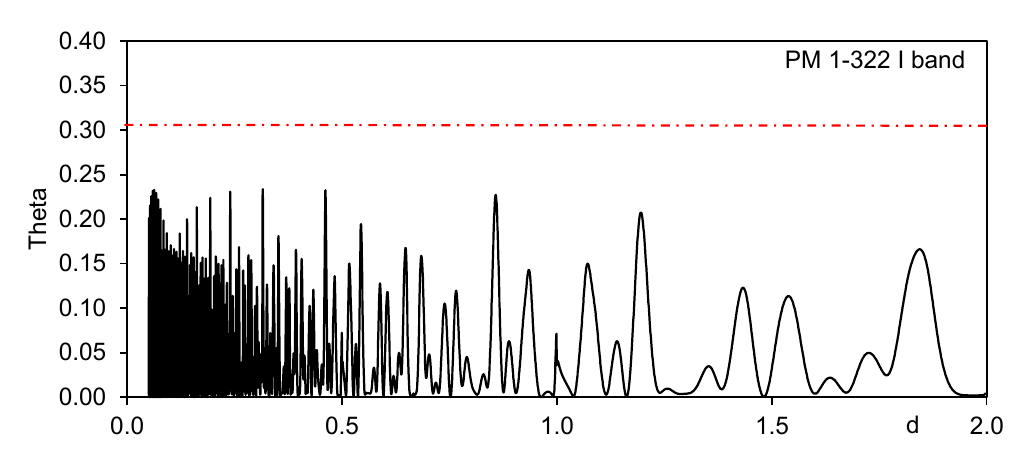}
\caption{Periodograms based on the analyses of (from top to bottom) ROAD $B$, $V$, $R{_C}$, and $I{_C}$ data with the generalised Lomb Scargle algorithm. The red dashed lines indicate an FAP of 1\,\%.}
\label{fig:periodograms_ROAD_data}
\end{center}
\end{figure}

To examine the ROAD observations for periodic variability, the Generalised Lomb Scargle algorithm from the programme package \textsc{PERANSO} \citep{PERANSO} was used. After removing a slight linear trend, the $B$, $V$, $R{_C}$, and $I{_C}$ data were examined for signals with a false alarm probability (FAP) less than 1\,\% (FAP < 0.01) in the period range of 0.05$-$2\,d. No significant periodic signal was found in the examined data sets (Figure \ref{fig:periodograms_ROAD_data}). While the number and pattern of peaks below the significance threshold suggest that there is some form of short-time variability in the period range of several tenths of a day, it is irregular or semi-regular at best.

\section{Interpretation and discussion} 
\label{interp}

In this section, we discuss several possible scenarios that might qualitatively explain the behaviour of our target object. They may or may not be correct; only future observations will be able to shed more light on this interesting system.

\subsection{Binary scenarios}

The two major and the one minor brightening events seen at roughly regular intervals in the IR light curves are suggestive of a periodic behaviour which could be associated with a binary nature of PM~1-322. In the following, we assume that the IR excess is caused by dust and not by the emission lines. Although we regularly refer to the hot and cold components, we caution that they do not necessarily have to be stars. For example, a hot component might be a pseudo-photosphere
or a hot disk surrounding an even hotter star. The cold component might be a dust cloud or a disk harbouring a cooler star. Bearing this in mind, one could speculate along the following lines.

Scenario (A): Forward and backward scattering on dust.
This scenario assumes a binary with a hot and a cold component. The cold component has, or is associated with, a dust cloud, which emits thermal radiation and also scatters the radiation from the star. The orbital inclination is close to an edge-on configuration.
Supposing the orbital period is 12 yr, one would observe major brightening episodes due to the forward scattering on an optically thin dust every 12 yr, which happens when the dust cloud is in front of the hot component \citep{budaj11}. The eclipse might occur at this moment. Apart from that, a minor brightening due to the backward scattering would be observed when the cloud is behind the hot component. These events would be more pronounced at shorter wavelengths and this is exactly what is observed in the W2, W1, and ZTF $r$ filters.
Unfortunately, this model fails to explain the opposite behaviour in the visible region. Furthermore, the observed amplitude
of about one magnitude in the W1 filter seems too high to be explained by scattering. Consequently, we think this scenario is likely not a satisfactory explanation.

Scenario (B): Reflection effect on dust.
Similar to model A, this scenario also assumes a hot and a cold component.
The cold component and/or the associated dust cloud is at least partially opaque and heated by the hot component on its `day' side. Assuming an orbital period of 6 yr, we would observe brightenings due to the thermal and scattered radiation from the cloud. This would be analogous to the reflection effect in binary systems and a brightening would be observed when the dust cloud is behind the hot component.
Again, this is neither suited to explain the anti-phase behaviour in the optical region nor the eclipse event which should occur at minimum IR light. What we see at shorter wavelengths looks like eclipses correlated with the brightenings in the IR.
This leads us to the following scenario.

Scenario (C): Dusty tails.
This scenario assumes a hot component and a cold dusty component, with the dust cloud eclipsing the hot component every 12 yr. The minor dimming event in the $V$ filter at MJD\,57200 is then a secondary eclipse. However, this requires the assumption that the hot component has a non-negligible size (or a disk) to be able to cause secondary eclipses.
We still need to explain the brightenings at IR wavelengths. Assuming the dust cloud has the shape of a tail trailing behind the cold component, its cross-section (projection) would be largest
near the primary eclipse causing the IR brightening. This would be analogous to the ellipsoidal variability of eclipsing binaries but with eclipses occurring at maxima rather than minima.
Six years later, when the tail is behind the hot component, we would observe another maximum in the IR, which is approximately what is observed. However, based on this model, one would expect that the minor and major IR brightenings are of similar strength or that the minor brightening is stronger due to the above-mentioned reflection effect. This, however, is not the case. It also remains hard to understand the brightening in ZTF $r$ before the eclipse and the little dips in the ZTF $g$ filter during the quiescent phase (MJD\,58200$-$59200). Consequently, it gets increasingly complicated and this scenario is likely not the correct explanation either.

Scenario (D): Dust clouds on an elliptical orbit.
In this scenario, we envisage a situation similar to scenario B (reflection effect), with the difference that the secondary component and its dust cloud are on an elliptical orbit with a period of about 6 yr.
When the cloud is at periastron, it becomes heated, which produces a brightening in the IR that is stronger in W1 than in W2. The probability that the dust cloud eclipses the hot component is highest at periastron, which could explain the observed dimmings at shorter wavelengths during the times of maxima in the IR. For this scenario to work, the dust cloud needs to be at least partially transparent. To explain the anti-phase behaviour and activity during the quiescent phase at MJD 58200$-$59200, one could envisage multiple dust clouds on similar elliptical orbits. This would be similar to the model of disintegrating asteroids \citep{neslusan17} proposed to explain the behaviour of Boyajian's star \citep{boyajian16}. In this object, minor dimming events are also observed before a major one. Dust embedded asteroids were also found orbiting white dwarfs \citep{vanderburg15,vanderburg20,vanderbosch20,manser19,gansicke19}. 
The problem with this scenario is the brightening before the eclipse in the ZTF $r$ filter at about 6300\,\AA, which would require temperatures well above 2000\,K and the dust would not survive at such temperatures.
One might speculate that if the secondary component is a giant star embedded in a dusty shell, the shell would sublimate at periastron, unveiling the spectrum of a giant star followed by brightenings in the ZTF $r$ filter.

\subsection{Puffed-up dusty disk scenario}
\label{subsec:puffed_up_disk}

Scenario (E): There may be another more simple explanation of the observed behaviour.
Let us assume a central hot star surrounded by an inner hotter gaseous disk and an outer cooler dusty disk. We assume that both disks are nearly edge-on. If the disks were to expand and get hotter for some reason, an increase of brightness in the IR and ZTF $r$ filters would be observed. At the same time, the puffed-up edges of the outer disk might eclipse the central star or the hotter inner disk, which leads to the dimming events in the optical region. One would expect that the dimming is stronger at shorter wavelengths due to dust extinction being stronger here -- and this is exactly what is observed.
This model also agrees with the spectroscopic data and the double-peaked line profiles. The major eclipse is still likely due to a dust enshrouded secondary star but, theoretically, even this could be understood by puffing-up the disk.

However, the origin of the disk remains to be explained, as well as the reason why it would expand and get hotter. It is interesting to note that, before doing so, there is a relatively short eruption-like event, which was observed in the ZTF data (MJD\,59270) and might trigger these changes. It may be a sort of eruption throwing out material that subsequently cools down and forms new dust. It is also conceivable that the hot star has a cool companion and the eruption takes actually place on the companion star. The material from this cool star is then transferred to the accretion disk increasing the mass accretion rate and thereby heating and puffing-up the disk. This would be analogous to what is observed in a symbiotic star when an expanded and flared disk builds up during active phases \citep{skopal05}.   
The inner disk may completely hide a central hot white dwarf from our sight. However, in the perpendicular direction,
the hot star is unobscured, ionising the circumstellar medium and giving rise to the strong emission lines.
Small variability in the disk may also explain the little dimming events in the ZTF $g$ filter during the quiescent period (MJD 58200$-$59200). The observed shift in the radial velocities of 5 km/s might somehow be associated with the companion.

Alternatively, there might be smaller bodies, their debris, inhomogeneities, or dust clouds in the disk that cause the major eclipse. Smaller variability in ZTF $r$ and $g$ before the eclipse or during the quiescent period may also easily be due to variability in the emission lines. The eruption-like events that preceded the eclipse may conceivably be due to small-body collision cascades or `a nebular reflection' of and eruption that took place at the hot star, which radiatively excited the material in the nebula that subsequently `shone' in the emission lines. In summary, we consider scenarios D \& E, or a combination thereof, as the most likely explanation for the observed kind of variability.

\section{Conclusion}

During the search for anti-phase variability at different wavelengths, we discovered that ZTFJ201451.59+120353.4 is an object with very peculiar variability properties. The ZTF $r$ and $g$ data show a one-magnitude deep, eclipse-like event with a duration of about half a year that occurred in 2022. Furthermore, the variability is characterised by dimming events in the optical region that are accompanied by brightening events in the red and IR regions. Apart from that, two fast eruption-like events were recorded in the ZTF $r$ data. Archival data from the WISE mission indicate long-term variability with a possible period of about 12 or 6 yr. Our follow-up photometric observations revealed a stochastic short-term variability with an amplitude of about 0.1\,mag on the timescale of about one hour. 

Most intriguing, ZTFJ201451.59+120353.4 identifies with the planetary nebula candidate PM~1-322. Its spectral energy distribution peaks in the mid-infrared region. Our high-resolution spectroscopic observations show strong, narrow, and mostly symmetric forbidden emission lines from highly ionised species and broader symmetric, double-peaked emission in H$\alpha$, which is very different from what is seen in earlier spectra obtained in 2007. Radial velocities derived from these lines are also different and shifted by about 5 km/s with respect to the values measured in 2007.

We speculate about the nature of the observed variability pattern. Many possible scenarios seem suitable to at least partly explain the observations and cannot be excluded. The forbidden emission lines from highly ionised species indicate the presence of a hot compact star embedded in an extended nebula, while the double-peaked emission in H$\alpha$ suggests a gaseous disk. The significant IR excess in the SED indicates the presence of dust. The observed eclipses and variability dictate that the gaseous disk, the dust, and a possible companion are in a nearly edge-on configuration.

While we prefer a scenario involving a puffed-up dusty disk (Section \ref{subsec:puffed_up_disk}), which readily agrees with the spectroscopic properties and is also able to explain the observed anti-phase variability, we stress that these are only speculations. Many open questions remain and it is clear that this most interesting system requires further long-term monitoring as well as UV and IR spectroscopic observations, which we herewith encourage. 

\begin{acknowledgements}
The authors would like to thank Dr. Augustin Skopal for his comments on the manuscript and Andrii Maliuk for help with the archival data.

  This work has made use of data from the European Space Agency (ESA) mission {\it Gaia} (\url{https://www.cosmos.esa.int/gaia}), processed by the {\it Gaia} Data Processing and Analysis Consortium (DPAC, \url{https://www.cosmos.esa.int/web/gaia/dpac/consortium}).  Funding for the DPAC has been provided by national institutions, in particular, the institutions participating in the {\it Gaia} Multilateral Agreement. This research has made use of the SIMBAD database, operated at CDS, Strasbourg, France.
  
  This publication makes use of data products from the Wide-field Infrared Survey Explorer, which is a joint project of the University of California, Los Angeles, and the Jet Propulsion Laboratory/California Institute of Technology, funded by the National Aeronautics and Space Administration.
  
  This publication also makes use of data products from NEOWISE, which is a project of the Jet Propulsion Laboratory/California Institute of Technology, funded by the Planetary Science Division of the National Aeronautics and Space Administration.
  
  Based on observations obtained with the Samuel Oschin Telescope 48-inch and the 60-inch Telescope 
at the Palomar Observatory as part of the Zwicky Transient Facility project. 
ZTF is supported by the National Science Foundation under Grants No. AST-1440341 and AST-2034437 and a collaboration including current partners Caltech, IPAC, the Weizmann Institute for Science, 
the Oskar Klein Center at Stockholm University, the University of Maryland, 
Deutsches Elektronen-Synchrotron and Humboldt University, the TANGO Consortium of Taiwan, 
the University of Wisconsin at Milwaukee, Trinity College Dublin, Lawrence Livermore National Laboratories, 
IN2P3, University of Warwick, Ruhr University Bochum, Northwestern University and former partners 
the University of Washington, Los Alamos National Laboratories, and Lawrence Berkeley National Laboratories. 
Operations are conducted by COO, IPAC, and UW.

Based on observations made with the Nordic Optical Telescope, owned in collaboration by the University of Turku and Aarhus University, and operated jointly by Aarhus University, the University of Turku and the University of Oslo, representing Denmark, Finland and Norway, the University of Iceland and Stockholm University at the Observatorio del Roque de los Muchachos, La Palma, Spain, of the Instituto de Astrof\'sica de Canarias.

This publication makes use of VOSA, 
developed under the Spanish Virtual Observatory project supported 
from the Spanish MICINN through grant AyA2008-02156.
This research has made use of the Spanish Virtual Observatory (https://svo.cab.inta-csic.es) project funded by MCIN/AEI/10.13039/501100011033/ through grant PID2020-112949GB-I00.

JB was supported by the VEGA 2/0031/22 and APVV-20-148 grants. This work was supported by the Erasmus+ programme of the European Union under
grant number 2020-1-CZ01-KA203-078200

\end{acknowledgements}

\bibliographystyle{aa}
\bibliography{main,budaj}

\begin{thebibliography}{55}
\expandafter\ifx\csname natexlab\endcsname\relax\def\natexlab#1{#1}\fi

\bibitem[{{Ahn} {et~al.}(2012){Ahn}, {Alexandroff}, {Allende Prieto},
  {Anderson}, {Anderton}, {Andrews}, {Aubourg}, {Bailey}, {Balbinot}, {Barnes},
  {Bautista}, {Beers}, {Beifiori}, {Berlind}, {Bhardwaj}, {Bizyaev}, {Blake},
  {Blanton}, {Blomqvist}, {Bochanski}, {Bolton}, {Borde}, {Bovy}, {Brandt},
  {Brinkmann}, {Brown}, {Brownstein}, {Bundy}, {Busca}, {Carithers}, {Carnero},
  {Carr}, {Casetti-Dinescu}, {Chen}, {Chiappini}, {Comparat}, {Connolly},
  {Crepp}, {Cristiani}, {Croft}, {Cuesta}, {da Costa}, {Davenport}, {Dawson},
  {de Putter}, {De Lee}, {Delubac}, {Dhital}, {Ealet}, {Ebelke}, {Edmondson},
  {Eisenstein}, {Escoffier}, {Esposito}, {Evans}, {Fan}, {Femen{\'\i}a
  Castell{\'a}}, {Fern{\'a}ndez Alvar}, {Ferreira}, {Filiz Ak}, {Finley},
  {Fleming}, {Font-Ribera}, {Frinchaboy}, {Garc{\'\i}a-Hern{\'a}ndez},
  {Garc{\'\i}a P{\'e}rez}, {Ge}, {G{\'e}nova-Santos}, {Gillespie}, {Girardi},
  {Gonz{\'a}lez Hern{\'a}ndez}, {Grebel}, {Gunn}, {Guo}, {Haggard}, {Hamilton},
  {Harris}, {Hawley}, {Hearty}, {Ho}, {Hogg}, {Holtzman}, {Honscheid},
  {Huehnerhoff}, {Ivans}, {Ivezi{\'c}}, {Jacobson}, {Jiang}, {Johansson},
  {Johnson}, {Kauffmann}, {Kirkby}, {Kirkpatrick}, {Klaene}, {Knapp}, {Kneib},
  {Le Goff}, {Leauthaud}, {Lee}, {Lee}, {Long}, {Loomis}, {Lucatello},
  {Lundgren}, {Lupton}, {Ma}, {Ma}, {MacDonald}, {Mack}, {Mahadevan}, {Maia},
  {Majewski}, {Makler}, {Malanushenko}, {Malanushenko}, {Manchado},
  {Mandelbaum}, {Manera}, {Maraston}, {Margala}, {Martell}, {McBride},
  {McGreer}, {McMahon}, {M{\'e}nard}, {Meszaros}, {Miralda-Escud{\'e}},
  {Montero-Dorta}, {Montesano}, {Morrison}, {Muna}, {Munn}, {Murayama},
  {Myers}, {Neto}, {Nguyen}, {Nichol}, {Nidever}, {Noterdaeme}, {Nuza},
  {Ogando}, {Olmstead}, {Oravetz}, {Owen}, {Padmanabhan},
  {Palanque-Delabrouille}, {Pan}, {Parejko}, {Parihar}, {P{\^a}ris},
  {Pattarakijwanich}, {Pepper}, {Percival}, {P{\'e}rez-Fournon},
  {P{\'e}rez-R{\`a}fols}, {Petitjean}, {Pforr}, {Pieri}, {Pinsonneault}, {Porto
  de Mello}, {Prada}, {Price-Whelan}, {Raddick}, {Rebolo}, {Rich}, {Richards},
  {Robin}, {Rocha-Pinto}, {Rockosi}, {Roe}, {Ross}, {Ross}, {Rossi},
  {Rubi{\~n}o-Martin}, {Samushia}, {Sanchez Almeida}, {S{\'a}nchez},
  {Santiago}, {Sayres}, {Schlegel}, {Schlesinger}, {Schmidt}, {Schneider},
  {Schultheis}, {Schwope}, {Sc{\'o}ccola}, {Seljak}, {Sheldon}, {Shen}, {Shu},
  {Simmerer}, {Simmons}, {Skibba}, {Skrutskie}, {Slosar}, {Sobreira}, {Sobeck},
  {Stassun}, {Steele}, {Steinmetz}, {Strauss}, {Streblyanska}, {Suzuki},
  {Swanson}, {Tal}, {Thakar}, {Thomas}, {Thompson}, {Tinker}, {Tojeiro},
  {Tremonti}, {Vargas Maga{\~n}a}, {Verde}, {Viel}, {Vikas}, {Vogt}, {Wake},
  {Wang}, {Weaver}, {Weinberg}, {Weiner}, {West}, {White}, {Wilson},
  {Wisniewski}, {Wood-Vasey}, {Yanny}, {Y{\`e}che}, {York}, {Zamora},
  {Zasowski}, {Zehavi}, {Zhao}, {Zheng}, {Zhu}, \&
  {Zinn}}]{2012ApJS..203...21A}
{Ahn}, C.~P., {Alexandroff}, R., {Allende Prieto}, C., {et~al.} 2012, \apjs,
  203, 21

\bibitem[{{Akras} {et~al.}(2019){Akras}, {Guzman-Ramirez}, {Leal-Ferreira}, \&
  {Ramos-Larios}}]{2019ApJS..240...21A}
{Akras}, S., {Guzman-Ramirez}, L., {Leal-Ferreira}, M.~L., \& {Ramos-Larios},
  G. 2019, \apjs, 240, 21

\bibitem[{{Allen}(1982)}]{allen82}
{Allen}, D.~A. 1982, in Astrophysics and Space Science Library, Vol.~95, IAU
  Colloq. 70: The Nature of Symbiotic Stars, ed. M.~{Friedjung} \& R.~{Viotti},
  27--42

\bibitem[{{Bailer-Jones} {et~al.}(2021){Bailer-Jones}, {Rybizki}, {Fouesneau},
  {Demleitner}, \& {Andrae}}]{2021AJ....161..147B}
{Bailer-Jones}, C.~A.~L., {Rybizki}, J., {Fouesneau}, M., {Demleitner}, M., \&
  {Andrae}, R. 2021, \aj, 161, 147

\bibitem[{{Bailer-Jones} {et~al.}(2018){Bailer-Jones}, {Rybizki}, {Fouesneau},
  {Mantelet}, \& {Andrae}}]{2018AJ....156...58B}
{Bailer-Jones}, C.~A.~L., {Rybizki}, J., {Fouesneau}, M., {Mantelet}, G., \&
  {Andrae}, R. 2018, \aj, 156, 58

\bibitem[{{Bayo} {et~al.}(2008){Bayo}, {Rodrigo}, {Barrado Y Navascu{\'e}s},
  {Solano}, {Guti{\'e}rrez}, {Morales-Calder{\'o}n}, \& {Allard}}]{bayo08}
{Bayo}, A., {Rodrigo}, C., {Barrado Y Navascu{\'e}s}, D., {et~al.} 2008, \aap,
  492, 277

\bibitem[{{Bellm} {et~al.}(2019){Bellm}, {Kulkarni}, {Graham}, {Dekany},
  {Smith}, {Riddle}, {Masci}, {Helou}, {Prince}, {Adams}, {Barbarino},
  {Barlow}, {Bauer}, {Beck}, {Belicki}, {Biswas}, {Blagorodnova}, {Bodewits},
  {Bolin}, {Brinnel}, {Brooke}, {Bue}, {Bulla}, {Burruss}, {Cenko}, {Chang},
  {Connolly}, {Coughlin}, {Cromer}, {Cunningham}, {De}, {Delacroix}, {Desai},
  {Duev}, {Eadie}, {Farnham}, {Feeney}, {Feindt}, {Flynn}, {Franckowiak},
  {Frederick}, {Fremling}, {Gal-Yam}, {Gezari}, {Giomi}, {Goldstein},
  {Golkhou}, {Goobar}, {Groom}, {Hacopians}, {Hale}, {Henning}, {Ho}, {Hover},
  {Howell}, {Hung}, {Huppenkothen}, {Imel}, {Ip}, {Ivezi{\'c}}, {Jackson},
  {Jones}, {Juric}, {Kasliwal}, {Kaspi}, {Kaye}, {Kelley}, {Kowalski},
  {Kramer}, {Kupfer}, {Landry}, {Laher}, {Lee}, {Lin}, {Lin}, {Lunnan},
  {Giomi}, {Mahabal}, {Mao}, {Miller}, {Monkewitz}, {Murphy}, {Ngeow},
  {Nordin}, {Nugent}, {Ofek}, {Patterson}, {Penprase}, {Porter}, {Rauch},
  {Rebbapragada}, {Reiley}, {Rigault}, {Rodriguez}, {van Roestel}, {Rusholme},
  {van Santen}, {Schulze}, {Shupe}, {Singer}, {Soumagnac}, {Stein}, {Surace},
  {Sollerman}, {Szkody}, {Taddia}, {Terek}, {Van Sistine}, {van Velzen},
  {Vestrand}, {Walters}, {Ward}, {Ye}, {Yu}, {Yan}, \& {Zolkower}}]{bellm19}
{Bellm}, E.~C., {Kulkarni}, S.~R., {Graham}, M.~J., {et~al.} 2019, \pasp, 131,
  018002

\bibitem[{{Boyajian} {et~al.}(2016){Boyajian}, {LaCourse}, {Rappaport},
  {Fabrycky}, {Fischer}, {Gandolfi}, {Kennedy}, {Korhonen}, {Liu}, {Moor},
  {Olah}, {Vida}, {Wyatt}, {Best}, {Brewer}, {Ciesla}, {Cs{\'a}k}, {Deeg},
  {Dupuy}, {Handler}, {Heng}, {Howell}, {Ishikawa}, {Kov{\'a}cs}, {Kozakis},
  {Kriskovics}, {Lehtinen}, {Lintott}, {Lynn}, {Nespral}, {Nikbakhsh},
  {Schawinski}, {Schmitt}, {Smith}, {Szabo}, {Szabo}, {Viuho}, {Wang},
  {Weiksnar}, {Bosch}, {Connors}, {Goodman}, {Green}, {Hoekstra}, {Jebson},
  {Jek}, {Omohundro}, {Schwengeler}, \& {Szewczyk}}]{boyajian16}
{Boyajian}, T.~S., {LaCourse}, D.~M., {Rappaport}, S.~A., {et~al.} 2016,
  \mnras, 457, 3988

\bibitem[{{Bressan} {et~al.}(2012){Bressan}, {Marigo}, {Girardi}, {Salasnich},
  {Dal Cero}, {Rubele}, \& {Nanni}}]{2012MNRAS.427..127B}
{Bressan}, A., {Marigo}, P., {Girardi}, L., {et~al.} 2012, \mnras, 427, 127

\bibitem[{{Budaj}(2011)}]{budaj11}
{Budaj}, J. 2011, \aap, 532, L12

\bibitem[{{Chen} {et~al.}(2020){Chen}, {Wang}, {Deng}, {de Grijs}, {Yang}, \&
  {Tian}}]{2020ApJS..249...18C}
{Chen}, X., {Wang}, S., {Deng}, L., {et~al.} 2020, \apjs, 249, 18

\bibitem[{{Djupvik} \& {Andersen}(2010)}]{2010ASSP...14..211D}
{Djupvik}, A.~A. \& {Andersen}, J. 2010, in Astrophysics and Space Science
  Proceedings, Vol.~14, Highlights of Spanish Astrophysics V, 211

\bibitem[{{Faltov{\'a}} {et~al.}(2021){Faltov{\'a}}, {Kallov{\'a}},
  {Pri{\v{s}}egen}, {Stan{\v{e}}k}, {Sup{\'\i}kov{\'a}}, {Xia}, {Bernhard},
  {H{\"u}mmerich}, \& {Paunzen}}]{2021A&A...656A.125F}
{Faltov{\'a}}, N., {Kallov{\'a}}, K., {Pri{\v{s}}egen}, M., {et~al.} 2021,
  \aap, 656, A125

\bibitem[{{Frew} \& {Parker}(2010)}]{2010PASA...27..129F}
{Frew}, D.~J. \& {Parker}, Q.~A. 2010, \pasa, 27, 129

\bibitem[{{Gaia Collaboration} {et~al.}(2018{\natexlab{a}}){Gaia
  Collaboration}, {Babusiaux}, {van Leeuwen}, {Barstow}, {Jordi}, {Vallenari},
  {Bossini}, {Bressan}, {Cantat-Gaudin}, {van Leeuwen}, {Brown}, {Prusti}, {de
  Bruijne}, {Bailer-Jones}, {Biermann}, {Evans}, {Eyer}, {Jansen}, {Klioner},
  {Lammers}, {Lindegren}, {Luri}, {Mignard}, {Panem}, {Pourbaix}, {Randich},
  {Sartoretti}, {Siddiqui}, {Soubiran}, {Walton}, {Arenou}, {Bastian},
  {Cropper}, {Drimmel}, {Katz}, {Lattanzi}, {Bakker}, {Cacciari},
  {Casta{\~n}eda}, {Chaoul}, {Cheek}, {De Angeli}, {Fabricius}, {Guerra},
  {Holl}, {Masana}, {Messineo}, {Mowlavi}, {Nienartowicz}, {Panuzzo},
  {Portell}, {Riello}, {Seabroke}, {Tanga}, {Th{\'e}venin}, {Gracia-Abril},
  {Comoretto}, {Garcia-Reinaldos}, {Teyssier}, {Altmann}, {Andrae}, {Audard},
  {Bellas-Velidis}, {Benson}, {Berthier}, {Blomme}, {Burgess}, {Busso},
  {Carry}, {Cellino}, {Clementini}, {Clotet}, {Creevey}, {Davidson}, {De
  Ridder}, {Delchambre}, {Dell'Oro}, {Ducourant},
  {Fern{\'a}ndez-Hern{\'a}ndez}, {Fouesneau}, {Fr{\'e}mat}, {Galluccio},
  {Garc{\'\i}a-Torres}, {Gonz{\'a}lez-N{\'u}{\~n}ez}, {Gonz{\'a}lez-Vidal},
  {Gosset}, {Guy}, {Halbwachs}, {Hambly}, {Harrison}, {Hern{\'a}ndez},
  {Hestroffer}, {Hodgkin}, {Hutton}, {Jasniewicz}, {Jean-Antoine-Piccolo},
  {Jordan}, {Korn}, {Krone-Martins}, {Lanzafame}, {Lebzelter}, {L{\"o}ffler},
  {Manteiga}, {Marrese}, {Mart{\'\i}n-Fleitas}, {Moitinho}, {Mora}, {Muinonen},
  {Osinde}, {Pancino}, {Pauwels}, {Petit}, {Recio-Blanco}, {Richards},
  {Rimoldini}, {Robin}, {Sarro}, {Siopis}, {Smith}, {Sozzetti}, {S{\"u}veges},
  {Torra}, {van Reeven}, {Abbas}, {Abreu Aramburu}, {Accart}, {Aerts},
  {Altavilla}, {{\'A}lvarez}, {Alvarez}, {Alves}, {Anderson}, {Andrei},
  {Anglada Varela}, {Antiche}, {Antoja}, {Arcay}, {Astraatmadja}, {Bach},
  {Baker}, {Balaguer-N{\'u}{\~n}ez}, {Balm}, {Barache}, {Barata}, {Barbato},
  {Barblan}, {Barklem}, {Barrado}, {Barros}, {Bartholom{\'e} Mu{\~n}oz},
  {Bassilana}, {Becciani}, {Bellazzini}, {Berihuete}, {Bertone}, {Bianchi},
  {Bienaym{\'e}}, {Blanco-Cuaresma}, {Boch}, {Boeche}, {Bombrun}, {Borrachero},
  {Bouquillon}, {Bourda}, {Bragaglia}, {Bramante}, {Breddels}, {Brouillet},
  {Br{\"u}semeister}, {Brugaletta}, {Bucciarelli}, {Burlacu}, {Busonero},
  {Butkevich}, {Buzzi}, {Caffau}, {Cancelliere}, {Cannizzaro}, {Carballo},
  {Carlucci}, {Carrasco}, {Casamiquela}, {Castellani}, {Castro-Ginard},
  {Charlot}, {Chemin}, {Chiavassa}, {Cocozza}, {Costigan}, {Cowell}, {Crifo},
  {Crosta}, {Crowley}, {Cuypers}, {Dafonte}, {Damerdji}, {Dapergolas}, {David},
  {David}, {de Laverny}, {De Luise}, {De March}, {de Martino}, {de Souza}, {de
  Torres}, {Debosscher}, {del Pozo}, {Delbo}, {Delgado}, {Delgado}, {Diakite},
  {Diener}, {Distefano}, {Dolding}, {Drazinos}, {Dur{\'a}n}, {Edvardsson},
  {Enke}, {Eriksson}, {Esquej}, {Eynard Bontemps}, {Fabre}, {Fabrizio},
  {Faigler}, {Falc{\~a}o}, {Farr{\`a}s Casas}, {Federici}, {Fedorets},
  {Fernique}, {Figueras}, {Filippi}, {Findeisen}, {Fonti}, {Fraile}, {Fraser},
  {Fr{\'e}zouls}, {Gai}, {Galleti}, {Garabato}, {Garc{\'\i}a-Sedano},
  {Garofalo}, {Garralda}, {Gavel}, {Gavras}, {Gerssen}, {Geyer}, {Giacobbe},
  {Gilmore}, {Girona}, {Giuffrida}, {Glass}, {Gomes}, {Granvik}, {Gueguen},
  {Guerrier}, {Guiraud}, {Guti{\'e}}, {Haigron}, {Hatzidimitriou}, {Hauser},
  {Haywood}, {Heiter}, {Helmi}, {Heu}, {Hilger}, {Hobbs}, {Hofmann}, {Holland},
  {Huckle}, {Hypki}, {Icardi}, {Jan{\ss}en}, {Jevardat de Fombelle}, {Jonker},
  {Juh{\'a}sz}, {Julbe}, {Karampelas}, {Kewley}, {Klar}, {Kochoska}, {Kohley},
  {Kolenberg}, {Kontizas}, {Kontizas}, {Koposov}, {Kordopatis},
  {Kostrzewa-Rutkowska}, {Koubsky}, {Lambert}, {Lanza}, {Lasne}, {Lavigne}, {Le
  Fustec}, {Le Poncin-Lafitte}, {Lebreton}, {Leccia}, {Leclerc},
  {Lecoeur-Taibi}, {Lenhardt}, {Leroux}, {Liao}, {Licata}, {Lindstr{\o}m},
  {Lister}, {Livanou}, {Lobel}, {L{\'o}pez}, {Managau}, {Mann}, {Mantelet},
  {Marchal}, {Marchant}, {Marconi}, {Marinoni}, {Marschalk{\'o}}, {Marshall},
  {Martino}, {Marton}, {Mary}, {Massari}, {Matijevi{\v{c}}}, {Mazeh},
  {McMillan}, {Messina}, {Michalik}, {Millar}, {Molina}, {Molinaro},
  {Moln{\'a}r}, {Montegriffo}, {Mor}, {Morbidelli}, {Morel}, {Morris},
  {Mulone}, {Muraveva}, {Musella}, {Nelemans}, {Nicastro}, {Noval},
  {O'Mullane}, {Ord{\'e}novic}, {Ord{\'o}{\~n}ez-Blanco}, {Osborne}, {Pagani},
  {Pagano}, {Pailler}, {Palacin}, {Palaversa}, {Panahi}, {Pawlak},
  {Piersimoni}, {Pineau}, {Plachy}, {Plum}, {Poggio}, {Poujoulet},
  {Pr{\v{s}}a}, {Pulone}, {Racero}, {Ragaini}, {Rambaux}, {Ramos-Lerate},
  {Regibo}, {Reyl{\'e}}, {Riclet}, {Ripepi}, {Riva}, {Rivard}, {Rixon},
  {Roegiers}, {Roelens}, {Romero-G{\'o}mez}, {Rowell}, {Royer}, {Ruiz-Dern},
  {Sadowski}, {Sagrist{\`a} Sell{\'e}s}, {Sahlmann}, {Salgado}, {Salguero},
  {Sanna}, {Santana-Ros}, {Sarasso}, {Savietto}, {Schultheis}, {Sciacca},
  {Segol}, {Segovia}, {S{\'e}gransan}, {Shih}, {Siltala}, {Silva}, {Smart},
  {Smith}, {Solano}, {Solitro}, {Sordo}, {Soria Nieto}, {Souchay}, {Spagna},
  {Spoto}, {Stampa}, {Steele}, {Steidelm{\"u}ller}, {Stephenson}, {Stoev},
  {Suess}, {Surdej}, {Szabados}, {Szegedi-Elek}, {Tapiador}, {Taris}, {Tauran},
  {Taylor}, {Teixeira}, {Terrett}, {Teyssandier}, {Thuillot}, {Titarenko},
  {Torra Clotet}, {Turon}, {Ulla}, {Utrilla}, {Uzzi}, {Vaillant}, {Valentini},
  {Valette}, {van Elteren}, {Van Hemelryck}, {Vaschetto}, {Vecchiato},
  {Veljanoski}, {Viala}, {Vicente}, {Vogt}, {von Essen}, {Voss}, {Votruba},
  {Voutsinas}, {Walmsley}, {Weiler}, {Wertz}, {Wevers}, {Wyrzykowski},
  {Yoldas}, {{\v{Z}}erjal}, {Ziaeepour}, {Zorec}, {Zschocke}, {Zucker},
  {Zurbach}, \& {Zwitter}}]{2018A&A...616A..10G}
{Gaia Collaboration}, {Babusiaux}, C., {van Leeuwen}, F., {et~al.}
  2018{\natexlab{a}}, \aap, 616, A10

\bibitem[{{Gaia Collaboration} {et~al.}(2018{\natexlab{b}}){Gaia
  Collaboration}, {Brown}, {Vallenari}, {Prusti}, {de Bruijne}, {Babusiaux},
  {Bailer-Jones}, {Biermann}, {Evans}, {Eyer}, {Jansen}, {Jordi}, {Klioner},
  {Lammers}, {Lindegren}, {Luri}, {Mignard}, {Panem}, {Pourbaix}, {Randich},
  {Sartoretti}, {Siddiqui}, {Soubiran}, {van Leeuwen}, {Walton}, {Arenou},
  {Bastian}, {Cropper}, {Drimmel}, {Katz}, {Lattanzi}, {Bakker}, {Cacciari},
  {Casta{\~n}eda}, {Chaoul}, {Cheek}, {De Angeli}, {Fabricius}, {Guerra},
  {Holl}, {Masana}, {Messineo}, {Mowlavi}, {Nienartowicz}, {Panuzzo},
  {Portell}, {Riello}, {Seabroke}, {Tanga}, {Th{\'e}venin}, {Gracia-Abril},
  {Comoretto}, {Garcia-Reinaldos}, {Teyssier}, {Altmann}, {Andrae}, {Audard},
  {Bellas-Velidis}, {Benson}, {Berthier}, {Blomme}, {Burgess}, {Busso},
  {Carry}, {Cellino}, {Clementini}, {Clotet}, {Creevey}, {Davidson}, {De
  Ridder}, {Delchambre}, {Dell'Oro}, {Ducourant},
  {Fern{\'a}ndez-Hern{\'a}ndez}, {Fouesneau}, {Fr{\'e}mat}, {Galluccio},
  {Garc{\'\i}a-Torres}, {Gonz{\'a}lez-N{\'u}{\~n}ez}, {Gonz{\'a}lez-Vidal},
  {Gosset}, {Guy}, {Halbwachs}, {Hambly}, {Harrison}, {Hern{\'a}ndez},
  {Hestroffer}, {Hodgkin}, {Hutton}, {Jasniewicz}, {Jean-Antoine-Piccolo},
  {Jordan}, {Korn}, {Krone-Martins}, {Lanzafame}, {Lebzelter}, {L{\"o}ffler},
  {Manteiga}, {Marrese}, {Mart{\'\i}n-Fleitas}, {Moitinho}, {Mora}, {Muinonen},
  {Osinde}, {Pancino}, {Pauwels}, {Petit}, {Recio-Blanco}, {Richards},
  {Rimoldini}, {Robin}, {Sarro}, {Siopis}, {Smith}, {Sozzetti}, {S{\"u}veges},
  {Torra}, {van Reeven}, {Abbas}, {Abreu Aramburu}, {Accart}, {Aerts},
  {Altavilla}, {{\'A}lvarez}, {Alvarez}, {Alves}, {Anderson}, {Andrei},
  {Anglada Varela}, {Antiche}, {Antoja}, {Arcay}, {Astraatmadja}, {Bach},
  {Baker}, {Balaguer-N{\'u}{\~n}ez}, {Balm}, {Barache}, {Barata}, {Barbato},
  {Barblan}, {Barklem}, {Barrado}, {Barros}, {Barstow}, {Bartholom{\'e}
  Mu{\~n}oz}, {Bassilana}, {Becciani}, {Bellazzini}, {Berihuete}, {Bertone},
  {Bianchi}, {Bienaym{\'e}}, {Blanco-Cuaresma}, {Boch}, {Boeche}, {Bombrun},
  {Borrachero}, {Bossini}, {Bouquillon}, {Bourda}, {Bragaglia}, {Bramante},
  {Breddels}, {Bressan}, {Brouillet}, {Br{\"u}semeister}, {Brugaletta},
  {Bucciarelli}, {Burlacu}, {Busonero}, {Butkevich}, {Buzzi}, {Caffau},
  {Cancelliere}, {Cannizzaro}, {Cantat-Gaudin}, {Carballo}, {Carlucci},
  {Carrasco}, {Casamiquela}, {Castellani}, {Castro-Ginard}, {Charlot},
  {Chemin}, {Chiavassa}, {Cocozza}, {Costigan}, {Cowell}, {Crifo}, {Crosta},
  {Crowley}, {Cuypers}, {Dafonte}, {Damerdji}, {Dapergolas}, {David}, {David},
  {de Laverny}, {De Luise}, {De March}, {de Martino}, {de Souza}, {de Torres},
  {Debosscher}, {del Pozo}, {Delbo}, {Delgado}, {Delgado}, {Di Matteo},
  {Diakite}, {Diener}, {Distefano}, {Dolding}, {Drazinos}, {Dur{\'a}n},
  {Edvardsson}, {Enke}, {Eriksson}, {Esquej}, {Eynard Bontemps}, {Fabre},
  {Fabrizio}, {Faigler}, {Falc{\~a}o}, {Farr{\`a}s Casas}, {Federici},
  {Fedorets}, {Fernique}, {Figueras}, {Filippi}, {Findeisen}, {Fonti},
  {Fraile}, {Fraser}, {Fr{\'e}zouls}, {Gai}, {Galleti}, {Garabato},
  {Garc{\'\i}a-Sedano}, {Garofalo}, {Garralda}, {Gavel}, {Gavras}, {Gerssen},
  {Geyer}, {Giacobbe}, {Gilmore}, {Girona}, {Giuffrida}, {Glass}, {Gomes},
  {Granvik}, {Gueguen}, {Guerrier}, {Guiraud}, {Guti{\'e}rrez-S{\'a}nchez},
  {Haigron}, {Hatzidimitriou}, {Hauser}, {Haywood}, {Heiter}, {Helmi}, {Heu},
  {Hilger}, {Hobbs}, {Hofmann}, {Holland}, {Huckle}, {Hypki}, {Icardi},
  {Jan{\ss}en}, {Jevardat de Fombelle}, {Jonker}, {Juh{\'a}sz}, {Julbe},
  {Karampelas}, {Kewley}, {Klar}, {Kochoska}, {Kohley}, {Kolenberg},
  {Kontizas}, {Kontizas}, {Koposov}, {Kordopatis}, {Kostrzewa-Rutkowska},
  {Koubsky}, {Lambert}, {Lanza}, {Lasne}, {Lavigne}, {Le Fustec}, {Le
  Poncin-Lafitte}, {Lebreton}, {Leccia}, {Leclerc}, {Lecoeur-Taibi},
  {Lenhardt}, {Leroux}, {Liao}, {Licata}, {Lindstr{\o}m}, {Lister}, {Livanou},
  {Lobel}, {L{\'o}pez}, {Managau}, {Mann}, {Mantelet}, {Marchal}, {Marchant},
  {Marconi}, {Marinoni}, {Marschalk{\'o}}, {Marshall}, {Martino}, {Marton},
  {Mary}, {Massari}, {Matijevi{\v{c}}}, {Mazeh}, {McMillan}, {Messina},
  {Michalik}, {Millar}, {Molina}, {Molinaro}, {Moln{\'a}r}, {Montegriffo},
  {Mor}, {Morbidelli}, {Morel}, {Morris}, {Mulone}, {Muraveva}, {Musella},
  {Nelemans}, {Nicastro}, {Noval}, {O'Mullane}, {Ord{\'e}novic},
  {Ord{\'o}{\~n}ez-Blanco}, {Osborne}, {Pagani}, {Pagano}, {Pailler},
  {Palacin}, {Palaversa}, {Panahi}, {Pawlak}, {Piersimoni}, {Pineau}, {Plachy},
  {Plum}, {Poggio}, {Poujoulet}, {Pr{\v{s}}a}, {Pulone}, {Racero}, {Ragaini},
  {Rambaux}, {Ramos-Lerate}, {Regibo}, {Reyl{\'e}}, {Riclet}, {Ripepi}, {Riva},
  {Rivard}, {Rixon}, {Roegiers}, {Roelens}, {Romero-G{\'o}mez}, {Rowell},
  {Royer}, {Ruiz-Dern}, {Sadowski}, {Sagrist{\`a} Sell{\'e}s}, {Sahlmann},
  {Salgado}, {Salguero}, {Sanna}, {Santana-Ros}, {Sarasso}, {Savietto},
  {Schultheis}, {Sciacca}, {Segol}, {Segovia}, {S{\'e}gransan}, {Shih},
  {Siltala}, {Silva}, {Smart}, {Smith}, {Solano}, {Solitro}, {Sordo}, {Soria
  Nieto}, {Souchay}, {Spagna}, {Spoto}, {Stampa}, {Steele},
  {Steidelm{\"u}ller}, {Stephenson}, {Stoev}, {Suess}, {Surdej}, {Szabados},
  {Szegedi-Elek}, {Tapiador}, {Taris}, {Tauran}, {Taylor}, {Teixeira},
  {Terrett}, {Teyssandier}, {Thuillot}, {Titarenko}, {Torra Clotet}, {Turon},
  {Ulla}, {Utrilla}, {Uzzi}, {Vaillant}, {Valentini}, {Valette}, {van Elteren},
  {Van Hemelryck}, {van Leeuwen}, {Vaschetto}, {Vecchiato}, {Veljanoski},
  {Viala}, {Vicente}, {Vogt}, {von Essen}, {Voss}, {Votruba}, {Voutsinas},
  {Walmsley}, {Weiler}, {Wertz}, {Wevers}, {Wyrzykowski}, {Yoldas},
  {{\v{Z}}erjal}, {Ziaeepour}, {Zorec}, {Zschocke}, {Zucker}, {Zurbach}, \&
  {Zwitter}}]{2018A&A...616A...1G}
{Gaia Collaboration}, {Brown}, A.~G.~A., {Vallenari}, A., {et~al.}
  2018{\natexlab{b}}, \aap, 616, A1

\bibitem[{{Gaia Collaboration} {et~al.}(2021){Gaia Collaboration}, {Brown},
  {Vallenari}, {Prusti}, {de Bruijne}, {Babusiaux}, {Biermann}, {Creevey},
  {Evans}, {Eyer}, {Hutton}, {Jansen}, {Jordi}, {Klioner}, {Lammers},
  {Lindegren}, {Luri}, {Mignard}, {Panem}, {Pourbaix}, {Randich}, {Sartoretti},
  {Soubiran}, {Walton}, {Arenou}, {Bailer-Jones}, {Bastian}, {Cropper},
  {Drimmel}, {Katz}, {Lattanzi}, {van Leeuwen}, {Bakker}, {Cacciari},
  {Casta{\~n}eda}, {De Angeli}, {Ducourant}, {Fabricius}, {Fouesneau},
  {Fr{\'e}mat}, {Guerra}, {Guerrier}, {Guiraud}, {Jean-Antoine Piccolo},
  {Masana}, {Messineo}, {Mowlavi}, {Nicolas}, {Nienartowicz}, {Pailler},
  {Panuzzo}, {Riclet}, {Roux}, {Seabroke}, {Sordo}, {Tanga}, {Th{\'e}venin},
  {Gracia-Abril}, {Portell}, {Teyssier}, {Altmann}, {Andrae}, {Bellas-Velidis},
  {Benson}, {Berthier}, {Blomme}, {Brugaletta}, {Burgess}, {Busso}, {Carry},
  {Cellino}, {Cheek}, {Clementini}, {Damerdji}, {Davidson}, {Delchambre},
  {Dell'Oro}, {Fern{\'a}ndez-Hern{\'a}ndez}, {Galluccio}, {Garc{\'\i}a-Lario},
  {Garcia-Reinaldos}, {Gonz{\'a}lez-N{\'u}{\~n}ez}, {Gosset}, {Haigron},
  {Halbwachs}, {Hambly}, {Harrison}, {Hatzidimitriou}, {Heiter},
  {Hern{\'a}ndez}, {Hestroffer}, {Hodgkin}, {Holl}, {Jan{\ss}en}, {Jevardat de
  Fombelle}, {Jordan}, {Krone-Martins}, {Lanzafame}, {L{\"o}ffler}, {Lorca},
  {Manteiga}, {Marchal}, {Marrese}, {Moitinho}, {Mora}, {Muinonen}, {Osborne},
  {Pancino}, {Pauwels}, {Petit}, {Recio-Blanco}, {Richards}, {Riello},
  {Rimoldini}, {Robin}, {Roegiers}, {Rybizki}, {Sarro}, {Siopis}, {Smith},
  {Sozzetti}, {Ulla}, {Utrilla}, {van Leeuwen}, {van Reeven}, {Abbas}, {Abreu
  Aramburu}, {Accart}, {Aerts}, {Aguado}, {Ajaj}, {Altavilla}, {{\'A}lvarez},
  {{\'A}lvarez Cid-Fuentes}, {Alves}, {Anderson}, {Anglada Varela}, {Antoja},
  {Audard}, {Baines}, {Baker}, {Balaguer-N{\'u}{\~n}ez}, {Balbinot}, {Balog},
  {Barache}, {Barbato}, {Barros}, {Barstow}, {Bartolom{\'e}}, {Bassilana},
  {Bauchet}, {Baudesson-Stella}, {Becciani}, {Bellazzini}, {Bernet}, {Bertone},
  {Bianchi}, {Blanco-Cuaresma}, {Boch}, {Bombrun}, {Bossini}, {Bouquillon},
  {Bragaglia}, {Bramante}, {Breedt}, {Bressan}, {Brouillet}, {Bucciarelli},
  {Burlacu}, {Busonero}, {Butkevich}, {Buzzi}, {Caffau}, {Cancelliere},
  {C{\'a}novas}, {Cantat-Gaudin}, {Carballo}, {Carlucci}, {Carnerero},
  {Carrasco}, {Casamiquela}, {Castellani}, {Castro-Ginard}, {Castro Sampol},
  {Chaoul}, {Charlot}, {Chemin}, {Chiavassa}, {Cioni}, {Comoretto}, {Cooper},
  {Cornez}, {Cowell}, {Crifo}, {Crosta}, {Crowley}, {Dafonte}, {Dapergolas},
  {David}, {David}, {de Laverny}, {De Luise}, {De March}, {De Ridder}, {de
  Souza}, {de Teodoro}, {de Torres}, {del Peloso}, {del Pozo}, {Delbo},
  {Delgado}, {Delgado}, {Delisle}, {Di Matteo}, {Diakite}, {Diener},
  {Distefano}, {Dolding}, {Eappachen}, {Edvardsson}, {Enke}, {Esquej}, {Fabre},
  {Fabrizio}, {Faigler}, {Fedorets}, {Fernique}, {Fienga}, {Figueras},
  {Fouron}, {Fragkoudi}, {Fraile}, {Franke}, {Gai}, {Garabato},
  {Garcia-Gutierrez}, {Garc{\'\i}a-Torres}, {Garofalo}, {Gavras}, {Gerlach},
  {Geyer}, {Giacobbe}, {Gilmore}, {Girona}, {Giuffrida}, {Gomel}, {Gomez},
  {Gonzalez-Santamaria}, {Gonz{\'a}lez-Vidal}, {Granvik},
  {Guti{\'e}rrez-S{\'a}nchez}, {Guy}, {Hauser}, {Haywood}, {Helmi}, {Hidalgo},
  {Hilger}, {H{\l}adczuk}, {Hobbs}, {Holland}, {Huckle}, {Jasniewicz},
  {Jonker}, {Juaristi Campillo}, {Julbe}, {Karbevska}, {Kervella}, {Khanna},
  {Kochoska}, {Kontizas}, {Kordopatis}, {Korn}, {Kostrzewa-Rutkowska},
  {Kruszy{\'n}ska}, {Lambert}, {Lanza}, {Lasne}, {Le Campion}, {Le Fustec},
  {Lebreton}, {Lebzelter}, {Leccia}, {Leclerc}, {Lecoeur-Taibi}, {Liao},
  {Licata}, {Lindstr{\o}m}, {Lister}, {Livanou}, {Lobel}, {Madrero Pardo},
  {Managau}, {Mann}, {Marchant}, {Marconi}, {Marcos Santos}, {Marinoni},
  {Marocco}, {Marshall}, {Martin Polo}, {Mart{\'\i}n-Fleitas}, {Masip},
  {Massari}, {Mastrobuono-Battisti}, {Mazeh}, {McMillan}, {Messina},
  {Michalik}, {Millar}, {Mints}, {Molina}, {Molinaro}, {Moln{\'a}r},
  {Montegriffo}, {Mor}, {Morbidelli}, {Morel}, {Morris}, {Mulone}, {Munoz},
  {Muraveva}, {Murphy}, {Musella}, {Noval}, {Ord{\'e}novic}, {Orr{\`u}},
  {Osinde}, {Pagani}, {Pagano}, {Palaversa}, {Palicio}, {Panahi}, {Pawlak},
  {Pe{\~n}alosa Esteller}, {Penttil{\"a}}, {Piersimoni}, {Pineau}, {Plachy},
  {Plum}, {Poggio}, {Poretti}, {Poujoulet}, {Pr{\v{s}}a}, {Pulone}, {Racero},
  {Ragaini}, {Rainer}, {Raiteri}, {Rambaux}, {Ramos}, {Ramos-Lerate}, {Re
  Fiorentin}, {Regibo}, {Reyl{\'e}}, {Ripepi}, {Riva}, {Rixon}, {Robichon},
  {Robin}, {Roelens}, {Rohrbasser}, {Romero-G{\'o}mez}, {Rowell}, {Royer},
  {Rybicki}, {Sadowski}, {Sagrist{\`a} Sell{\'e}s}, {Sahlmann}, {Salgado},
  {Salguero}, {Samaras}, {Sanchez Gimenez}, {Sanna}, {Santove{\~n}a},
  {Sarasso}, {Schultheis}, {Sciacca}, {Segol}, {Segovia}, {S{\'e}gransan},
  {Semeux}, {Shahaf}, {Siddiqui}, {Siebert}, {Siltala}, {Slezak}, {Smart},
  {Solano}, {Solitro}, {Souami}, {Souchay}, {Spagna}, {Spoto}, {Steele},
  {Steidelm{\"u}ller}, {Stephenson}, {S{\"u}veges}, {Szabados}, {Szegedi-Elek},
  {Taris}, {Tauran}, {Taylor}, {Teixeira}, {Thuillot}, {Tonello}, {Torra},
  {Torra}, {Turon}, {Unger}, {Vaillant}, {van Dillen}, {Vanel}, {Vecchiato},
  {Viala}, {Vicente}, {Voutsinas}, {Weiler}, {Wevers}, {Wyrzykowski}, {Yoldas},
  {Yvard}, {Zhao}, {Zorec}, {Zucker}, {Zurbach}, \&
  {Zwitter}}]{2021A&A...649A...1G}
{Gaia Collaboration}, {Brown}, A.~G.~A., {Vallenari}, A., {et~al.} 2021, \aap,
  649, A1

\bibitem[{{G{\"a}nsicke} {et~al.}(2019){G{\"a}nsicke}, {Schreiber}, {Toloza},
  {Gentile Fusillo}, {Koester}, \& {Manser}}]{gansicke19}
{G{\"a}nsicke}, B.~T., {Schreiber}, M.~R., {Toloza}, O., {et~al.} 2019, \nat,
  576, 61

\bibitem[{{Gardner} {et~al.}(2006){Gardner}, {Mather}, {Clampin}, {Doyon},
  {Greenhouse}, {Hammel}, {Hutchings}, {Jakobsen}, {Lilly}, {Long}, {Lunine},
  {McCaughrean}, {Mountain}, {Nella}, {Rieke}, {Rieke}, {Rix}, {Smith},
  {Sonneborn}, {Stiavelli}, {Stockman}, {Windhorst}, \& {Wright}}]{jwst06}
{Gardner}, J.~P., {Mather}, J.~C., {Clampin}, M., {et~al.} 2006, \ssr, 123, 485

\bibitem[{{Green} {et~al.}(2019){Green}, {Schlafly}, {Zucker}, {Speagle}, \&
  {Finkbeiner}}]{2019ApJ...887...93G}
{Green}, G.~M., {Schlafly}, E., {Zucker}, C., {Speagle}, J.~S., \&
  {Finkbeiner}, D. 2019, \apj, 887, 93

\bibitem[{{Hambsch}(2012)}]{hambsch12}
{Hambsch}, F.~J. 2012, JAAVSO, 40, 1003

\bibitem[{{Hanuschik} {et~al.}(1988){Hanuschik}, {Kozok}, \&
  {Kaiser}}]{1988A&A...189..147H}
{Hanuschik}, R.~W., {Kozok}, J.~R., \& {Kaiser}, D. 1988, \aap, 189, 147

\bibitem[{{Jones} \& {Boffin}(2017)}]{jones17nat}
{Jones}, D. \& {Boffin}, H. M.~J. 2017, Nature Astronomy, 1, 0117

\bibitem[{{Jones} {et~al.}(2017){Jones}, {Van Winckel}, {Aller}, {Exter}, \&
  {De Marco}}]{jones17}
{Jones}, D., {Van Winckel}, H., {Aller}, A., {Exter}, K., \& {De Marco}, O.
  2017, \aap, 600, L9

\bibitem[{{Kenyon} {et~al.}(1988){Kenyon}, {Fernandez-Castro}, \&
  {Stencel}}]{kenyon88}
{Kenyon}, S.~J., {Fernandez-Castro}, T., \& {Stencel}, R.~E. 1988, \aj, 95,
  1817

\bibitem[{{Kochanek} {et~al.}(2017){Kochanek}, {Shappee}, {Stanek}, {Holoien},
  {Thompson}, {Prieto}, {Dong}, {Shields}, {Will}, {Britt}, {Perzanowski}, \&
  {Pojma{\'n}ski}}]{kochanek17}
{Kochanek}, C.~S., {Shappee}, B.~J., {Stanek}, K.~Z., {et~al.} 2017, \pasp,
  129, 104502

\bibitem[{{Kohoutek}(1982)}]{kohoutek82}
{Kohoutek}, L. 1982, Information Bulletin on Variable Stars, 2113, 1

\bibitem[{{Kwok}(2000)}]{kwok00}
{Kwok}, S. 2000, {The Origin and Evolution of Planetary Nebulae}

\bibitem[{{Lamers} {et~al.}(1998){Lamers}, {Zickgraf}, {de Winter}, {Houziaux},
  \& {Zorec}}]{lamers98}
{Lamers}, H. J.~G.~L.~M., {Zickgraf}, F.-J., {de Winter}, D., {Houziaux}, L.,
  \& {Zorec}, J. 1998, \aap, 340, 117

\bibitem[{{Lee} \& {Hyung}(2013)}]{lee13}
{Lee}, S.~J. \& {Hyung}, S. 2013, \aap, 549, A65

\bibitem[{{Mainzer} {et~al.}(2011){Mainzer}, {Bauer}, {Grav}, {Masiero},
  {Cutri}, {Dailey}, {Eisenhardt}, {McMillan}, {Wright}, {Walker}, {Jedicke},
  {Spahr}, {Tholen}, {Alles}, {Beck}, {Brandenburg}, {Conrow}, {Evans},
  {Fowler}, {Jarrett}, {Marsh}, {Masci}, {McCallon}, {Wheelock}, {Wittman},
  {Wyatt}, {DeBaun}, {Elliott}, {Elsbury}, {Gautier}, {Gomillion}, {Leisawitz},
  {Maleszewski}, {Micheli}, \& {Wilkins}}]{mainzer11}
{Mainzer}, A., {Bauer}, J., {Grav}, T., {et~al.} 2011, \apj, 731, 53

\bibitem[{{Manser} {et~al.}(2019){Manser}, {G{\"a}nsicke}, {Eggl}, {Hollands},
  {Izquierdo}, {Koester}, {Landstreet}, {Lyra}, {Marsh}, {Meru}, {Mustill},
  {Rodr{\'\i}guez-Gil}, {Toloza}, {Veras}, {Wilson}, {Burleigh}, {Davies},
  {Farihi}, {Gentile Fusillo}, {de Martino}, {Parsons}, {Quirrenbach}, {Raddi},
  {Reffert}, {Del Santo}, {Schreiber}, {Silvotti}, {Toonen}, {Villaver},
  {Wyatt}, {Xu}, \& {Portegies Zwart}}]{manser19}
{Manser}, C.~J., {G{\"a}nsicke}, B.~T., {Eggl}, S., {et~al.} 2019, Science,
  364, 66

\bibitem[{{Mendez} {et~al.}(1982){Mendez}, {Gathier}, \& {Niemela}}]{mendez82}
{Mendez}, R.~H., {Gathier}, R., \& {Niemela}, V.~S. 1982, \aap, 116, L5

\bibitem[{{Miranda} {et~al.}(2010){Miranda}, {V{\'a}zquez}, {Guerrero},
  {Pereira}, \& {I{\~n}iguez-Gar{\'\i}n}}]{2010PASA...27..199M}
{Miranda}, L.~F., {V{\'a}zquez}, R., {Guerrero}, M.~A., {Pereira}, C.~B., \&
  {I{\~n}iguez-Gar{\'\i}n}, E. 2010, \pasa, 27, 199

\bibitem[{{Miszalski} {et~al.}(2011){Miszalski}, {Miko{\l}ajewska},
  {K{\"o}ppen}, {Rauch}, {Acker}, {Cohen}, {Frew}, {Moffat}, {Parker}, {Jones},
  \& {Udalski}}]{miszalski11}
{Miszalski}, B., {Miko{\l}ajewska}, J., {K{\"o}ppen}, J., {et~al.} 2011, \aap,
  528, A39

\bibitem[{{Neslu{\v{s}}an} \& {Budaj}(2017)}]{neslusan17}
{Neslu{\v{s}}an}, L. \& {Budaj}, J. 2017, \aap, 600, A86

\bibitem[{{Paunzen} \& {Vanmunster}(2016)}]{PERANSO}
{Paunzen}, E. \& {Vanmunster}, T. 2016, Astronomische Nachrichten, 337, 239

\bibitem[{{Pereira} \& {Miranda}(2005)}]{2005A&A...433..579P}
{Pereira}, C.~B. \& {Miranda}, L.~F. 2005, \aap, 433, 579

\bibitem[{{Pollacco} {et~al.}(1992){Pollacco}, {Kilkenny}, {Marang}, {van Wyk},
  \& {Roberts}}]{pollacco92}
{Pollacco}, D.~L., {Kilkenny}, D., {Marang}, F., {van Wyk}, F., \& {Roberts},
  G. 1992, \mnras, 256, 669

\bibitem[{{Rodrigo} {et~al.}(2012){Rodrigo}, {Solano}, \& {Bayo}}]{rodrigo12}
{Rodrigo}, C., {Solano}, E., \& {Bayo}, A. 2012, {SVO Filter Profile Service
  Version 1.0}, IVOA Working Draft 15 October 2012

\bibitem[{{Saad} {et~al.}(2004){Saad}, {Kub{\'a}t}, {Koubsk{\'y}}, {Harmanec},
  {{\v{S}}koda}, {Kor{\v{c}}{\'a}kov{\'a}}, {Krti{\v{c}}ka}, {{\v{S}}lechta},
  {Bo{\v{z}}i{\'c}}, {Ak}, {Hadrava}, \& {Votruba}}]{2004A&A...419..607S}
{Saad}, S.~M., {Kub{\'a}t}, J., {Koubsk{\'y}}, P., {et~al.} 2004, \aap, 419,
  607

\bibitem[{{Schmid}(1989)}]{1989A&A...211L..31S}
{Schmid}, H.~M. 1989, \aap, 211, L31

\bibitem[{{Shappee} {et~al.}(2014){Shappee}, {Prieto}, {Grupe}, {Kochanek},
  {Stanek}, {De Rosa}, {Mathur}, {Zu}, {Peterson}, {Pogge}, {Komossa}, {Im},
  {Jencson}, {Holoien}, {Basu}, {Beacom}, {Szczygie{\l}}, {Brimacombe},
  {Adams}, {Campillay}, {Choi}, {Contreras}, {Dietrich}, {Dubberley},
  {Elphick}, {Foale}, {Giustini}, {Gonzalez}, {Hawkins}, {Howell}, {Hsiao},
  {Koss}, {Leighly}, {Morrell}, {Mudd}, {Mullins}, {Nugent}, {Parrent},
  {Phillips}, {Pojmanski}, {Rosing}, {Ross}, {Sand}, {Terndrup}, {Valenti},
  {Walker}, \& {Yoon}}]{shappe14}
{Shappee}, B.~J., {Prieto}, J.~L., {Grupe}, D., {et~al.} 2014, \apj, 788, 48

\bibitem[{{Skopal}(2005)}]{skopal05}
{Skopal}, A. 2005, \aap, 440, 995

\bibitem[{{Stone}(1977)}]{stone77}
{Stone}, R.~P.~S. 1977, \apj, 218, 767

\bibitem[{{Telting} {et~al.}(2014){Telting}, {Avila}, {Buchhave}, {Frandsen},
  {Gandolfi}, {Lindberg}, {Stempels}, {Prins}, \& {NOT
  staff}}]{2014AN....335...41T}
{Telting}, J.~H., {Avila}, G., {Buchhave}, L., {et~al.} 2014, Astronomische
  Nachrichten, 335, 41

\bibitem[{{van Winckel}(2003)}]{vanwinckel03}
{van Winckel}, H. 2003, \araa, 41, 391

\bibitem[{{Van Winckel} {et~al.}(2014){Van Winckel}, {Jorissen}, {Exter},
  {Raskin}, {Prins}, {Perez Padilla}, {Merges}, \& {Pessemier}}]{vanwinckel14}
{Van Winckel}, H., {Jorissen}, A., {Exter}, K., {et~al.} 2014, \aap, 563, L10

\bibitem[{{Vanderbosch} {et~al.}(2020){Vanderbosch}, {Hermes}, {Dennihy},
  {Dunlap}, {Izquierdo}, {Tremblay}, {Cho}, {G{\"a}nsicke}, {Toloza}, {Bell},
  {Montgomery}, \& {Winget}}]{vanderbosch20}
{Vanderbosch}, Z., {Hermes}, J.~J., {Dennihy}, E., {et~al.} 2020, \apj, 897,
  171

\bibitem[{{Vanderburg} {et~al.}(2015){Vanderburg}, {Johnson}, {Rappaport},
  {Bieryla}, {Irwin}, {Lewis}, {Kipping}, {Brown}, {Dufour}, {Ciardi}, {Angus},
  {Schaefer}, {Latham}, {Charbonneau}, {Beichman}, {Eastman}, {McCrady},
  {Wittenmyer}, \& {Wright}}]{vanderburg15}
{Vanderburg}, A., {Johnson}, J.~A., {Rappaport}, S., {et~al.} 2015, \nat, 526,
  546

\bibitem[{{Vanderburg} {et~al.}(2020){Vanderburg}, {Rappaport}, {Xu},
  {Crossfield}, {Becker}, {Gary}, {Murgas}, {Blouin}, {Kaye}, {Palle}, {Melis},
  {Morris}, {Kreidberg}, {Gorjian}, {Morley}, {Mann}, {Parviainen}, {Pearce},
  {Newton}, {Carrillo}, {Zuckerman}, {Nelson}, {Zeimann}, {Brown},
  {Tronsgaard}, {Klein}, {Ricker}, {Vanderspek}, {Latham}, {Seager}, {Winn},
  {Jenkins}, {Adams}, {Benneke}, {Berardo}, {Buchhave}, {Caldwell},
  {Christiansen}, {Collins}, {Col{\'o}n}, {Daylan}, {Doty}, {Doyle},
  {Dragomir}, {Dressing}, {Dufour}, {Fukui}, {Glidden}, {Guerrero}, {Guo},
  {Heng}, {Henriksen}, {Huang}, {Kaltenegger}, {Kane}, {Lewis}, {Lissauer},
  {Morales}, {Narita}, {Pepper}, {Rose}, {Smith}, {Stassun}, \&
  {Yu}}]{vanderburg20}
{Vanderburg}, A., {Rappaport}, S.~A., {Xu}, S., {et~al.} 2020, \nat, 585, 363

\bibitem[{{Wesson}(2016)}]{wesson16}
{Wesson}, R. 2016, \mnras, 456, 3774

\bibitem[{{Wilson}(1950)}]{wilson50}
{Wilson}, O.~C. 1950, \apj, 111, 279

\bibitem[{{Wright} {et~al.}(2010){Wright}, {Eisenhardt}, {Mainzer}, {Ressler},
  {Cutri}, {Jarrett}, {Kirkpatrick}, {Padgett}, {McMillan}, {Skrutskie},
  {Stanford}, {Cohen}, {Walker}, {Mather}, {Leisawitz}, {Gautier}, {McLean},
  {Benford}, {Lonsdale}, {Blain}, {Mendez}, {Irace}, {Duval}, {Liu}, {Royer},
  {Heinrichsen}, {Howard}, {Shannon}, {Kendall}, {Walsh}, {Larsen}, {Cardon},
  {Schick}, {Schwalm}, {Abid}, {Fabinsky}, {Naes}, \& {Tsai}}]{wright10}
{Wright}, E.~L., {Eisenhardt}, P. R.~M., {Mainzer}, A.~K., {et~al.} 2010, \aj,
  140, 1868

\bibitem[{{Zickgraf}(2003)}]{zickgraf03}
{Zickgraf}, F.~J. 2003, \aap, 408, 257

\end{thebibliography}

\end{document}